\documentclass[pra,preprint,superscriptaddress,floatfix,showpacs]{revtex4-1}

\usepackage{amsmath,amsfonts,amssymb}
\usepackage{graphicx}

\def\beq{\begin{equation}}
\def\eeq{\end{equation}}
\def\beqn{\begin{eqnarray}}
\def\eeqn{\end{eqnarray}}

\def\x {{\bf x}}
\def\y {{\bf y}}

\def\Q {{\bf Q}}

\def\X {{\bf X}}
\def\Y {{\bf Y}}

\begin{document}

\title{Solvable Model of a Generic Trapped Mixture of Interacting Bosons: 
Reduced Density Matrices and Proof of\\ Bose-Einstein Condensation}
\author{Ofir E. Alon}
\affiliation{Department of Physics, University of Haifa at Oranim, Tivon 36006, Israel}

\begin{abstract}
A mixture of two kinds of identical bosons, 
species $1$ with $N_1$ bosons of mass $m_1$ and 
species $2$ with $N_2$ bosons of mass $m_2$,
held in a harmonic potential of frequency $\omega$ 
and interacting by harmonic intra-species and inter-species particle-particle interactions 
of strengths $\lambda_1$, $\lambda_2$, and $\lambda_{12}$ is discussed. 
This is an exactly-solvable model of a generic mixture of trapped interacting bosons
which allows one to investigate and determine analytically properties of interest. 
For a start, closed form expressions for the frequencies, ground-state energy, and wave-function
of the mixture are obtained and briefly analyzed as a function of 
the masses, numbers of particles, and strengths of interactions. 
To prove Bose-Einstein condensation of the mixture three steps are needed.
First, we integrate the all-particle density matrix,
employing a four-parameter matrix-recurrence relations,
down to the lowest-order intra-species and inter-species reduced density matrices of the mixture.
Second, the coupled Gross-Pitaevskii (mean-field) equations of the mixture are solved analytically.
Third, we analyze the mixture's reduced density matrices in the limit of an infinite number of particles 
of both species $1$ and $2$
(when the interaction parameters, i.e., the products of the number of particles times the intra-species and inter-species
interaction strengths, are held fixed)
and prove that: 
(i) Both species $1$ and $2$ are 100\% condensed;
(ii) The inter-species reduced density matrix per particle is separable and
given by the product of the intra-species reduced density matrices per particle;
and
(iii) The mixture's energy per particle, and reduced density matrices and densities per particle
all coincide with the Gross-Pitaevskii quantities. 
Finally, when the infinite-particle limit is taken with respect to, say,
species $1$ only (with interaction parameters held fixed) we prove that:
(iv) Only species $1$ is 100\% condensed and its reduced density matrix and density per particle,
as well as the mixture's energy per particle, coincide with 
the Gross-Pitaevskii quantities of species $1$ alone; 
and
(v) The inter-species reduced density matrix per particle is nonetheless separable and
given by the product of the intra-species reduced density matrices per particle.
Implications are briefly discussed.
\end{abstract}


\maketitle 

\section{Introduction}\label{INTRO}

Mixtures of Bose-Einstein condensates made of ultra-cold quantum gases 
and their physical properties have attracted much attention in the past twenty years, 
see for example \cite{B1,B2,B3,B4,B5,B6,B7,B8,B9,B10,B11,B12,B13,B14,B15,B16,B17,B18,B19,B20,B21,B22,B23,B24,B25,B26}. 
The ground state of trapped mixtures has been studied both at the mean-field level (within Gross-Pitaevskii theory)
and by different many-body theoretical and numerical tools.
Unlike their older sibling -- Bose-Einstein condensates made of a single species -- 
for which there have been rigorous results
connecting 
in the infinite-particle limit 
the many-body and mean-field expressions for
the energy per particle, density per particle, 
and providing proof of $100\%$ Bose-Einstein condensation in the ground state
\cite{Yngvason_PRA,Lieb_PRL}, 
there are to the best of our knowledge no such results for trapped mixtures.

To address this topic, 
we present an exactly-solvable model of a generic mixture of trapped interacting bosons
which allows one, for a start, to compute analytically the energy and wave-function of the mixture
for any number of particles.
From there, and with some effort as we shall see below, 
intra-species and inter-species reduced density matrices are computed and analyzed.
This allows us to get concrete results in the infinite-particle limit
(when the products of the number of particles times the intra-species and inter-species
interaction strengths are held fixed)
on the energy per particle, intra-species and inter-species reduced density matrices per particle,
and to prove Bose-Einstein condensate of each of the species.
By solving analytically for the ground state at mean-field level (within Gross-Pitaevskii theory),
we are also able to compare in the infinite-particle limit 
the many-body and mean-field results.
For a mixture there is a nice twist on the infinite-particle limit,
which may be taken with respect to both species or with respect to one of them.
The conclusions from both infinite-particle-limit procedures are compared and contrasted.

The model we present for the mixture is that of two species of indistinguishable bosons, 
species $1$ with $N_1$ bosons of mass $m_1$ and 
species $2$ with $N_2$ bosons of mass $m_2$,
held in a harmonic potential of frequency $\omega$ 
and interacting by harmonic intra-species and inter-species particle-particle interactions 
of strengths $\lambda_1$, $\lambda_2$, and $\lambda_{12}$, respectively.
This is the harmonic-interaction model for trapped bosonic mixtures.
The harmonic-interaction model for identical particles has been widely used for
single-species bosons \cite{HIM_Cohen,HIM_Yan,HIM_Po1,HIM_Benchmarks,HIM_Po0,Schilling_FER_BOS,Ental_FER_BOS},
fermions \cite{HIM_Po0,Schilling_FER_BOS,Ental_FER_BOS,Schilling_FER_HIM_WORK,Axel_MCTDHF_HIM},
and un-trapped (i.e., translationally-invariant)
bosonic mixtures \cite{HIM_MIX_JPA1,HIM_MIX_JPA2,HIM_MIX_PRE}.
Most recently,
we were able to solve a specific case of the harmonic-interaction model for trapped mixtures,
a trapped symmetric mixture for which $N_1=N_2$, $m_1=m_2$, and $\lambda_1=\lambda_2$, 
and showed that the energy per particle and densities per particle 
(i.e., the diagonal of reduced density matrices per particle) 
converge in the infinite-particle limit to their mean-field analogs \cite{BB_HIM_SYM}.
To treat the general mixture with unequal numbers of particles, masses, and interaction strengths,
and especially to derive its reduced density matrices and prove Bose-Einstein condensation,
it is needed to generalize the analytical techniques developed 
by Cohen and Lee \cite{HIM_Cohen} for single-species bosons and extended in \cite{BB_HIM_SYM}
for the symmetric mixture substantially further,
which is done below. 

The structure of the paper is as follows. In Sec.~\ref{HIM_MIX}
we present the harmonic-interaction model for a generic trapped mixture of bosons and discuss
its ground-state energy and wave-function.
In Sec.~\ref{MIX_INF_BEC_PROOF} we construct explicitly the lowest-order intra-species and
inter-species reduced density matrices of the mixture,
solve analytically the two-coupled Gross-Pitaevskii equations,
perform the infinite-particle limit in which the many-body and mean-field results are compared,
and prove Bose-Einstein condensation.
Concluding remarks are provided in Sec.~\ref{CON_REM}.
Further details of the derivations are collected in the appendixes. 

\section{The Harmonic-Interaction Model for Trapped Bosonic Mixtures}\label{HIM_MIX}

Consider a mixture of two distinguishable 
types of identical bosons which we label $1$ and $2$.
The bosons are trapped in a three-dimensional isotropic harmonic potential of frequency $\omega$ 
and interact via harmonic particle-particle interactions.
In the present work we only deal with the ground state of the trapped mixture.
We treat the case of a generic mixture.
Namely,
a mixture consisting 
of $N_1$ bosons of type $1$ and mass $m_1$ 
and $N_2$ bosons of type $2$ and mass $m_2$.
The total number of particles is denoted by $N=N_1+N_2$.
Furthermore, the two intra-species interaction strengths are denoted by $\lambda_1$ and $\lambda_2$,
and the inter-species interaction strength by $\lambda_{12}$.
Positive values of $\lambda_1$, $\lambda_2$, and $\lambda_{12}$ mean 
attractive particle-particle interactions
whereas negative values imply repulsive interactions \cite{REMARK_INTERACTIONS}.

The Hamiltonian of the mixture is then given by ($\hbar=1$)
\beqn\label{HAM_MIX}
& & \hat H(\x_1,\ldots,\x_{N_1},\y_1,\ldots,\y_{N_2}) = \nonumber \\
& & \quad = \sum_{j=1}^{N_1} \left( -\frac{1}{2m_1} \frac{\partial^2}{\partial \x_j^2} + \frac{1}{2} m_1\omega_1^2 \x_j^2 \right) + 
 \sum_{j=1}^{N_2} \left( -\frac{1}{2m_2} \frac{\partial^2}{\partial \y_j^2} + \frac{1}{2} m_2\omega_2^2 \y_j^2 \right) + \nonumber \\
& & \quad + \lambda_1 \sum_{1 \le j < k}^{N_1} (\x_j-\x_k)^2 + 
\lambda_2 \sum_{1 \le j < k}^{N_2} (\y_j-\y_k)^2 +  
\lambda_{12} \sum_{j=1}^{N_1} \sum_{k=1}^{N_2} (\x_j-\y_k)^2. \
\eeqn
Here, the coordinates $\x_j$ denote bosons of type $1$ and $\y_k$ bosons of type $2$.
We work in Cartesian coordinates where the vector $\x=(x_1,x_2,x_3)$ denotes 
the position of a boson of type $1$
in three dimensions,
and $\frac{1}{i}\frac{\partial }{\partial \x}=
\frac{1}{i}\left(\frac{\partial}{\partial x_1},\frac{\partial}{\partial x_2},\frac{\partial}{\partial x_3}\right)$ its momentum.
To avoid cumbersome notation,
we denote $\x^2 \equiv x_1^2+x_2^2+x_3^2$ and
$\frac{\partial^2}{\partial \x^2} \equiv \frac{\partial^2}{\partial x_1^2} + \frac{\partial^2}{\partial x_2^2}    
+ \frac{\partial^2}{\partial x_3^2}$.
Analogous notation is employed for the vector $\y$ of a boson of type $2$.

To diagonalize the Hamiltonian (\ref{HAM_MIX}) we transform to the Jacoby coordinates \cite{HIM_MIX_PRE}
\beqn\label{MIX_COOR}
& & \Q_k = \frac{1}{\sqrt{k(k+1)}} \sum_{j=1}^{k} (\x_{k+1}-\x_j), \qquad 1 \le k \le N_1-1,  \nonumber \\
& & \Q_{N_1-1+k} = \frac{1}{\sqrt{k(k+1)}} \sum_{j=1}^{k} (\y_{k+1}-\y_j), \qquad 1 \le k \le N_2-1,  \nonumber \\
& & \Q_{N-1} = \sqrt{\frac{N_2}{N_1}} \sum_{j=1}^{N_1} \x_j - \sqrt{\frac{N_1}{N_2}} \sum_{j=1}^{N_2} \y_j, \nonumber \\
& & \Q_N = \frac{m_1}{M} \sum_{j=1}^{N_1} \x_j + \frac{m_2}{M} \sum_{j=1}^{N_2} \y_j. \
\eeqn
The meaning of (\ref{MIX_COOR}) is as follows: 
The first group of $N_1-1$ coordinates are relative coordinates of the bosons of species $1$;
the second group of $N_2-1$ coordinates are relative coordinates the bosons of species $2$;
$\Q_{N-1}$ can be seen as a relative coordinate between the center-of-mass of the species $1$ 
and the center-of-mass of the species $2$ bosons; 
and, finally, $\Q_N$ is the center-of-mass coordinate of all particles in the mixture.

Using (\ref{MIX_COOR}) and (\ref{MIX_KIN_1})-(\ref{MIX_KIN_2}) in Appendix \ref{APP_A}, 
the Hamiltonian (\ref{HAM_MIX}) transforms to the diagonal form
\beqn\label{HAM_DIAG}
& & \hat H(\Q_1,\ldots,\Q_N) = 
\sum_{k=1}^{N_1-1} \left( -\frac{1}{2m_1} \frac{\partial^2}{\partial \Q_k^2} + 
\frac{1}{2} m_1\Omega_1^2 \Q_k^2 \right) + \nonumber \\
& & \quad + \sum_{k=N_1}^{N-2} \left( -\frac{1}{2m_2} \frac{\partial^2}{\partial \Q_k^2} + 
\frac{1}{2} m_2\Omega_2^2 \Q_k^2 \right) + 
\left(-\frac{1}{2M_{12}} \frac{\partial^2}{\partial \Q_{N-1}^2} + \frac{1}{2} M_{12} \Omega_{12}^2 \Q_{N-1}^2\right) + \nonumber \\
& & \quad +\left(-\frac{1}{2M} \frac{\partial^2}{\partial \Q_N^2} + \frac{1}{2} M \omega^2 \Q_N^2\right), \qquad
M_{12} = \frac{m_1m_2}{M}, \ M=N_1m_1+N_2m_2. \
\eeqn
The transformed Hamiltonian of the mixture (\ref{HAM_DIAG}) is that of $N$ uncoupled harmonic oscillators
having the four masses $m_1$, $m_2$, $M_{12}$ (relative mass of the mixture), and $M$ (total mass of the mixture),
and four frequencies
\beqn\label{MIX_FREQ}
& & \Omega_1 = \sqrt{\omega^2 + \frac{2}{m_1}(N_1\lambda_1+N_2\lambda_{12})}, \qquad
 \Omega_2 = \sqrt{\omega^2 + \frac{2}{m_2}(N_2\lambda_2+N_1\lambda_{12})}, \nonumber \\
& & 
 \Omega_{12} =  \sqrt{\omega^2 + 2\left(\frac{N_1}{m_2} + \frac{N_2}{m_1}\right)\lambda_{12}} =
 \sqrt{\omega^2 + \frac{2\lambda_{12}}{M_{12}}}, \qquad
 \omega. \
\eeqn
The multiplicity of the frequencies is $N_1-1$, $N_2-1$, $1$, and $1$, respectively,
corresponding to the types and numbers of Jabobi coordinates \cite{REMARK_DEGEN}.

The frequencies (\ref{MIX_FREQ}) must be positive in order for a bound solution to exist.
This dictates bounds on both the intra-species $\lambda_1$ and $\lambda_2$ 
and inter-species $\lambda_{12}$ interactions 
which are:
\beqn\label{BOUNDS}
& & \Omega_{12}^2=\omega^2 + \frac{2\lambda_{12}}{M_{12}}>0 \quad \Longrightarrow \quad 
\lambda_{12} > -\frac{M_{12}\omega^2}{2}, \nonumber \\
& & \Omega_{1}^2 = \omega^2 + \frac{2}{m_1}(N_1\lambda_1+N_2\lambda_{12})>0
\quad \Longrightarrow \quad \lambda_1 > - m_1 \frac{N_2}{N_1} \lambda_{12} - \frac{m_1\omega^2}{2N_1}, \nonumber \\
& & \Omega_{2}^2 = \omega^2 + \frac{2}{m_2}(N_2\lambda_2+N_1\lambda_{12})>0
\quad \Longrightarrow \quad \lambda_2 > - m_2 \frac{N_1}{N_2} \lambda_{12} - \frac{m_2\omega^2}{2N_2}. \
\eeqn
The meaning of these bounds are as follows:
The inter-species interaction $\lambda_{12}$ is bounded from below
by the frequency of the trap and the relative mass,
irrespective of the intra-species interactions $\lambda_1$ and $\lambda_2$, 
otherwise the mixture cannot be trapped in the harmonic potential.
On the other hand, the intra-species interaction $\lambda_1$ is limited by the chosen 
inter-species interaction $\lambda_{12}$, the numbers of particles, and the mass $m_1$, 
and analogously $\lambda_2$.

We can now proceed and 
prescribe the normalized ground-state wave-function
\beqn\label{WAVE_FUN_1}
& & \Psi(\Q_1,\ldots,\Q_N) = 
\left(\frac{m_1\Omega_1}{\pi}\right)^{\frac{3(N_1-1)}{4}}
\left(\frac{m_2\Omega_2}{\pi}\right)^{\frac{3(N_2-1)}{4}}
\left(\frac{M_{12}\Omega_{12}}{\pi}\right)^{\frac{3}{4}}
\left(\frac{M\omega}{\pi}\right)^{\frac{3}{4}} \times \nonumber \\
& & \quad \times
e^{-\frac{1}{2} \left(m_1\Omega_1 \sum_{k=1}^{N_1-1} \Q_k^2 + m_2\Omega_2 \sum_{k=N_1}^{N-2} \Q_k^2 +
M_{12}\Omega_{12} \Q_{N-1}^2 + M\omega \Q_N^2 \right)}, \
\eeqn
along with the ground-state energy
\beqn\label{HIM_MIX_GS_E}
& & E = \frac{3}{2} \left[ (N_1-1) \Omega_1 + (N_2-1) \Omega_2 + \Omega_{12} + \omega \right] = 
 \frac{3}{2} \Bigg[(N_1-1) \sqrt{\omega^2 + \frac{2}{m_1}(N_1\lambda_1+N_2\lambda_{12})} + \nonumber \\
& & \quad + (N_2-1)\sqrt{\omega^2 + \frac{2}{m_2}(N_2\lambda_2+N_1\lambda_{12})} + \sqrt{\omega^2 + \frac{2\lambda_{12}}{M_{12}}} + \omega \Bigg] \
\eeqn 
of the trapped mixture \cite{REMARK_NO_TRAP}.
Using the bounds for $\lambda_1$, $\lambda_2$, and $\lambda_{12}$ in (\ref{BOUNDS}),
we obtain that the ground-state energy of the mixture 
is bound from below by $E > \frac{3}{2}\omega$,
which is obtained for $\Omega_{1} \to 0^+$, $\Omega_{2} \to 0^+$, and $\Omega_{12} \to 0^+$.
This means that all relative degrees of freedom are marginally bound,
and essentially only the center-of-mass degree of freedom is bound in the harmonic trap.
The system is then predominantly repulsive.
At the other end, when the mixture is predominantly attractive,
the energy is unbound from above.

To express the wave-function with respect to the original spatial coordinates we use relations 
(\ref{MIX_POT_1}) and find
\beqn\label{WAVE_FUN_2}
& & \Psi(\x_1,\ldots,\x_{N_1},\y_1,\ldots,\y_{N_2}) = \left(\frac{m_1\Omega_1}{\pi}\right)^{\frac{3(N_1-1)}{4}}
\left(\frac{m_2\Omega_2}{\pi}\right)^{\frac{3(N_2-1)}{4}}
\left(\frac{M_{12}\Omega_{12}}{\pi}\right)^{\frac{3}{4}}
\left(\frac{M\omega}{\pi}\right)^{\frac{3}{4}} \times \nonumber \\
& & \quad \times 
e^{-\frac{\alpha_1}{2} \sum_{j=1}^{N_1} \x_j^2 - \beta_1 \sum_{1 \le j < k}^{N_1} \x_j \cdot \x_k} \times
e^{-\frac{\alpha_2}{2} \sum_{j=1}^{N_2} \y_j^2 - \beta_2 \sum_{1 \le j < k}^{N_2} \y_j \cdot \y_k} \times
e^{+\gamma \sum_{j=1}^{N_1} \sum_{k=1}^{N_2} \x_j \cdot \y_k},  \
\eeqn 
where the parameters are
\beqn\label{WAVE_FUN_3}
& & \alpha_1 = m_1 \left[\Omega_1\left(1-\frac{1}{N_1}\right) 
+ (m_2N_2\Omega_{12} + m_1N_1\omega) \frac{1}{MN_1} \right] = m_1\Omega_1+\beta_1, \nonumber \\
& & \beta_1 = m_1 \left[- \Omega_1 \frac{1}{N_1} 
+ (m_2N_2\Omega_{12} + m_1N_1\omega) \frac{1}{MN_1} \right], \nonumber \\
& & \alpha_2 = m_2 \left[\Omega_2\left(1-\frac{1}{N_2}\right) 
+ (m_1N_1\Omega_{12} + m_2N_2\omega) \frac{1}{MN_2} \right] = m_2\Omega_2+\beta_2, \nonumber \\
& & \beta_2 = m_2 \left[- \Omega_2 \frac{1}{N_2} 
+ (m_1N_1\Omega_{12} + m_2N_2\omega) \frac{1}{MN_2} \right], \nonumber \\
& & \gamma=\frac{m_1m_2}{M}(\Omega_{12}-\omega) = M_{12}(\Omega_{12}-\omega). \
\eeqn
The ground-state wave-function of the mixture (\ref{WAVE_FUN_2}) is
seen to be comprised of a product of a type $1$ boson part, 
a type $2$ boson part, 
and a coupling $1$--$2$ part.
As required by indistinguishability of identical bosons,
$\Psi$ is symmetric to permutation of the coordinates $\x_j$ and $\x_k$ of any two $1$ type bosons,
and likewise symmetric to permutation of the coordinates $\y_j$ and $\y_k$ of any two $2$ type bosons,
but is not symmetric to permutation of the distinguishable $1$ and $2$ type bosons.

The harmonic-interaction model 
for a generic trapped mixture of interacting bosons presented above admits
a wealth of properties that can all be studied in principle analytically.
After all, the ground-state wave-function and energy are given as explicit simple functions
of all parameters  -- masses $m_1$ and $m_2$, numbers of particles $N_1$ and $N_2$, 
interactions strengths $\lambda_1$, $\lambda_2$ and $\lambda_{12}$, and trapping frequency $\omega$.
All quantum properties of the mixture can be computed from its wave-function,
although, as we shall see in the next section, not necessarily without some effort.

\section{Reduced Density Matrices and Proof of Bose-Einstein Condensation}\label{MIX_INF_BEC_PROOF}

To show that the individual species are Bose-Einstein condensed three steps are required.
The first, 
the computation of the intra-species reduced one-particle density matrices.
To discuss the complementary question of separability requires the inter-species reduced two-body density matrix.
These are calculated analytically in Sec.~\ref{RDMs_MIX} 
as a function of the masses, interaction strengths, and numbers of particles.
The second step is to solve, again analytically, the system at the mean-field level.
This means finding the solution to the two-species coupled Gross-Pitaevskii equations.
This is performed in Sec.~\ref{GP_BB_GEN}.
The third and final step is to perform the infinite-particle limit and to compare and contrast
the exact and mean-field solutions in this limit,
which is carried out in Sec.~\ref{INF_MB_GP_MIX}. 
In a mixture one can perform 
the infinite-particle limit with respect to the two species or 
with respect to one of the species. 
We will discuss and compare both infinite-particle-limit 
procedures below.

\subsection{The intra-species and inter-species reduced density matrices}\label{RDMs_MIX}

We start from the full $N$-particle density matrix of the mixture,
\beqn\label{MB_DENS}
& & \Psi(\x_1,\ldots,\x_{N_1},\y_1,\ldots,\y_{N_2})\Psi^\ast(\x'_1,\ldots,\x'_{N_1},\y'_1,\ldots,\y'_{N_2})  = \nonumber \\
& & \left(\frac{m_1\Omega_1}{\pi}\right)^{\frac{3(N_1-1)}{2}}
\left(\frac{m_2\Omega_2}{\pi}\right)^{\frac{3(N_2-1)}{2}}
\left(\frac{M_{12}\Omega_{12}}{\pi}\right)^{\frac{3}{2}}
\left(\frac{M\omega}{\pi}\right)^{\frac{3}{2}} \times \nonumber \\
& & \quad \times 
e^{-\frac{\alpha_1}{2} \sum_{j=1}^{N_1} (\x_j^2+{\x'_j}^2) - 
\beta_1 \sum_{1 \le j < k}^{N_1} (\x_j \cdot \x_k + \x'_j \cdot \x'_k)} \times \nonumber \\
& & \quad \times 
e^{-\frac{\alpha_2}{2} \sum_{j=1}^{N_2} (\y_j^2+{\y'_j}^2) - 
\beta_2 \sum_{1 \le j < k}^{N_2} (\y_j \cdot \y_k + \y'_j \cdot \y'_k)} \times \nonumber \\
& & \quad \times
e^{+\gamma \sum_{j=1}^{N_1} \sum_{k=1}^{N_2} (\x_j \cdot \y_k + \x'_j \cdot \y'_k)},  \
\eeqn 
which for convenience is here normalized to one,
$\int d\x_1\cdots d\x_{N_1}d\y_1\cdots d\y_{N_2}|\Psi|^2 =1$.
The intra-species reduced one-body density matrices of the mixture, 
\beqn\label{DENS_A_B}
& & \rho_1(\x,\x') =N_1 \int d\x_2 \cdots d\x_{N_1} d\y_1 \cdots d\y_{N_2} 
\Psi(\x,\x_2,\ldots,\x_{N_1},\y_1,\y_2,\ldots,\y_{N_2}) \times \nonumber \\
& & \qquad \qquad \qquad \times \Psi^\ast(\x',\x_2,\ldots,\x_{N_1},\y_1,\y_2,\ldots,\y_{N_2}), \nonumber \\
& & \rho_2(\y,\y') =N_2 \int d\x_1 \cdots d\x_{N_1} d\y_2 \cdots d\y_{N_2} 
\Psi(\x_1,\x_2,\ldots,\x_{N_1},\y,\y_2,\ldots,\y_{N_2}) \times \nonumber \\
& & \qquad \qquad \qquad \times \Psi^\ast(\x_1,\x_2,\ldots,\x_{N_1},\y',\y_2,\ldots,\y_{N_2}), \
\eeqn
are given by integrating over $N-1$ coordinates, 
and the inter-species reduced two-body density matrix 
\beqn\label{DENS_AB}
& & \rho_{12}(\x,\x',\y,\y') = N_1 N_2 \int d\x_2 \cdots d\x_{N_1} d\y_2 \cdots d\y_{N_2} 
\Psi(\x,\x_2,\ldots,\x_{N_1},\y,\y_2,\ldots,\y_{N_2}) \times \nonumber \\
& & \qquad \qquad \qquad \qquad \times \Psi^\ast(\x',\x_2,\ldots,\x_{N_1},\y',\y_2,\ldots,\y_{N_2}), \
\eeqn
which is the lowest-order 
inter-species reduced density matrix of the mixture,
is obtained by integrating over $N-2$ coordinates.

To obtain the lowest-order reduced density matrices
(\ref{DENS_A_B}) and (\ref{DENS_AB})
we need to perform {\it multiple} integrations of 
the $N$-particle
density matrix of the mixture (\ref{MB_DENS}).
Moreover, the latter contains a coupling $1$--$2$ part because of the 
inter-species interaction.
Thus, without an appropriate construction, 
the task becomes quickly impractical with increasing $N_1$ and $N_2$.
Below, to perform the integration, 
we derive a four-parameter matrix-recurrence relations, 
thus generalizing the one-parameter and two-parameter vector-recurrence relations
put forward, respectively, in the cases of the single-species \cite{HIM_Cohen} 
and symmetric-mixture \cite{BB_HIM_SYM}
harmonic-interaction models.

To integrate the $N$-particle density matrix
we begin by introducing the auxiliary function
\beqn\label{FUN_N_HALF_1}
& & F_{N_1,N_2}(\x_1,\ldots,\x_{N_1},\y_1,\ldots,\y_{N_2};\x'_1,\ldots,\x'_{N_1},\y'_1,\ldots,\y'_{N_2}; \nonumber \\
& & \quad \alpha_1,\beta_1,\alpha_2,\beta_2,C_{N_1,N_2},C'_{N_1,N_2},D_{N_1,N_2},D'_{N_1,N_2}) = \nonumber \\
& & \quad = e^{-\frac{\alpha_1}{2} \sum_{j=1}^{N_1} (\x_j^2+{\x'_j}^2) - 
\beta_1 \sum_{1 \le j < k}^{N_1} (\x_j \cdot \x_k + \x'_j \cdot \x'_k)} \,
e^{-\frac{\alpha_2}{2} \sum_{j=1}^{N_2} (\y_j^2+{\y'_j}^2) - 
\beta_2 \sum_{1 \le j < k}^{N_2} (\y_j \cdot \y_k + \y'_j \cdot \y'_k)} \times \nonumber \\
& & \quad \times
e^{-\frac{1}{4}C_{N_1,N_2} (\X_{N_1}+\X'_{N_1})^2}
e^{-\frac{1}{4}C'_{N_1,N_2} (\Y_{N_2}+\Y'_{N_2})^2} \times \nonumber \\
& & \quad \times e^{+\frac{1}{2}D_{N_1,N_2} (\X_{N_1}+\X'_{N_1})(\Y_{N_2}+\Y'_{N_2})}
e^{+\frac{1}{2}D'_{N_1,N_2} (\X_{N_1}-\X'_{N_1})(\Y_{N_2}-\Y'_{N_2})}, \
\eeqn  
where $\alpha_1$, $\beta_1$, $\alpha_2$, and $\beta_2$,
are given in (\ref{WAVE_FUN_3}),
\beq\label{FUN_N_HALF_2}
\X_{N_1} = \sum_{j=1}^{N_1} \x_j, \quad
\X'_{N_1} = \sum_{j=1}^{N_1} \x'_j, \quad
\Y_{N_2} = \sum_{j=1}^{N_2} \y_j, \quad
\Y'_{N_2} = \sum_{j=1}^{N_2} \y'_j
\eeq
are vectors \cite{REMARK_VECTORS}, 
and $C_{N_1,N_2}$, $C'_{N_1,N_2}$, $D_{N_1,N_2}$, and $D'_{N_1,N_2}$ are constants to play 
a further role below.
We note 
that if one were to equate the $N$-particle density (\ref{MB_DENS}) 
and the auxiliary function $F_{N_1,N_2}$ then this would imply 
that 
\beq\label{F_N1_N2_IN_CON}
C_{N_1,N_2}=0, \quad C'_{N_1,N_2}=0, \quad D_{N_1,N_2}=\gamma, \quad D'_{N_1,N_2}=\gamma.
\eeq 
We will perform the integration of the auxiliary function $F_{N_1,N_2}$ in two steps,
first by integrating the $\y_{N_2}, \y'_{N_2}=\y_{N_2}$ variables
and then the $\x_{N_1}, \x'_{N_1}=\x_{N_1}$ variables.
We shall call the first step the horizontal reduction of the auxiliary function $F_{N_1,N_2}$
and the second step the vertical reduction.
The sequences of integrations to be performed below may be written symbolically as
\beq\label{F_N1_N2_MATRIX_REDUCE}
\begin{array}{ccccc}
F_{N_1,N_2} & \Longrightarrow & F_{N_1,1} & \Rightarrow & F_{N_1,0} \\
\big\Downarrow & & \big\Downarrow & & \big\Downarrow \\
F_{1,N_2} & \Longrightarrow & F_{1,1} &  & F_{1,0} \\
\Downarrow & &  & &  \\
F_{0,N_2} & \Longrightarrow & F_{0,1} &  & \
\end{array} \ .
\eeq
The end terms of these integrations, $F_{1,1}$ and $F_{1,0}$, $F_{0,1}$, will be connected with the 
reduced density matrices (\ref{DENS_AB}) and (\ref{DENS_A_B}), respectively.

Thus, to perform the multiple integration steps in the horizontal reduction (\ref{F_N1_N2_MATRIX_REDUCE}),
we seek for a recurrence relation and write
\beqn\label{FUN_N_HALF_M1_1}
& & \int d\y_{N_2} F_{N_1,N_2} = \nonumber \\
& & = e^{-\frac{\alpha_1}{2} \sum_{j=1}^{N_1} (\x_j^2+{\x'_j}^2) - 
\beta_1 \sum_{1 \le j < k}^{N_1} (\x_j \cdot \x_k + \x'_j \cdot \x'_k)} \,
e^{-\frac{\alpha_2}{2} \sum_{j=1}^{N_2-1} (\y_j^2+{\y'_j}^2) - 
\beta_2 \sum_{1 \le j < k}^{N_2-1} (\y_j \cdot \y_k + \y'_j \cdot \y'_k)} \times \nonumber \\
& & \times
e^{-\frac{1}{4}C_{N_1,N_2} (\X_{N_1}+\X'_{N_1})^2}
e^{-\frac{1}{4}C'_{N_1,N_2} (\Y_{N_2-1}+\Y'_{N_2-1})^2} \times \nonumber \\
& & \times e^{+\frac{1}{2}D_{N_1,N_2} (\X_{N_1}+\X'_{N_1})(\Y_{N_2-1}+\Y'_{N_2-1})} 
e^{+\frac{1}{2}D'_{N_1,N_2} (\X_{N_1}-\X'_{N_1})(\Y_{N_2-1}-\Y'_{N_2-1})} \times  \nonumber \\
& & \times \int d\y_{N_2} e^{-(\alpha_2+C'_{N_1,N_2})\y_{N_2}^2 - 
[(\beta_2+C'_{N_1,N_2})(\Y_{N_2-1}+\Y'_{N_2-1}) - D_{N_1,N_2}(\X_{N_1}+\X'_{N_1})]\y_{N_2}} = \nonumber \\
& & = \left(\frac{\pi}{\alpha_2+C'_{N_1,N_2}}\right)^\frac{3}{2} 
 F_{N_1,N_2-1}(\x_1,\ldots,\x_{N_1},\y_1,\ldots,\y_{N_2-1};\x'_1,\ldots,\x'_{N_1},\y'_1,\ldots,\y'_{N_2-1}; \nonumber \\
& & \qquad \alpha_1,\beta_1,\alpha_2,\beta_2,C_{N_1,N_2-1},C'_{N_1,N_2-1},D_{N_1,N_2-1},D'_{N_1,N_2-1}), \ 
\eeqn
where the Gaussian integral
$\int d\y e^{- a \y^2 - 2b \y \cdot \X} = \left(\frac{\pi}{a}\right)^{\frac{3}{2}} e^{\frac{b^2\X^2}{a}}$
is used and $\int d\y_{N_2} F_{N_1,N_2} \ldots$ implicitly implies that $\y'_{N_2}=\y_{N_2}$
is used in the integrand.
The equality (\ref{FUN_N_HALF_M1_1}) relates the 
auxiliary function $F_{N_1,N_2}$ with $N_1+N_2$ coordinates and constants 
$C_{N_1,N_2}$, $C'_{N_1,N_2}$, $D_{N_1,N_2}$, and $D'_{N_1,N_2}$, 
to the auxiliary 
function $F_{N_1,N_2-1}$ of the same functional form,
with $N_1+(N_2-1)$ coordinates and corresponding constants 
$C_{N_1,N_2-1}$, $C'_{N_1,N_2-1}$, $D_{N_1,N_2-1}$, and $D'_{N_1,N_2-1}$
which depend on all constants appearing in $F_{N_1,N_2}$ and read
\beqn\label{FUN_N_HALF_M1_2}
& & C_{N_1,N_2-1} = C_{N_1,N_2} - \frac{D^2_{N_1,N_2}}{\alpha_2+C'_{N_1,N_2}}, \nonumber \\
& & C'_{N_1,N_2-1} = C'_{N_1,N_2} - \frac{(\beta_2+C'_{N_1,N_2})^2}{\alpha_2+C'_{N_1,N_2}}, \nonumber \\
& & D_{N_1,N_2-1} = D_{N_1,N_2} - \frac{\beta_2+C'_{N_1,N_2}}{\alpha_2+C'_{N_1,N_2}}D_{N_1,N_2}, \nonumber \\
& & D'_{N_1,N_2-1} = D'_{N_1,N_2}.
\eeqn 
Recall that the initial values
of $C_{N_1,N_2}$, $C'_{N_1,N_2}$, $D_{N_1,N_2}$, and $D'_{N_1,N_2}$
are given by (\ref{F_N1_N2_IN_CON})
and 
obtained
when we equate the 
auxiliary function $F_{N_1,N_2}$ and the $N$-body density (\ref{MB_DENS}).

According to (\ref{F_N1_N2_MATRIX_REDUCE}),
the horizontal reduction makes a `stopover' at the auxiliary function
\beqn\label{FUN_N_HALF_END}
& & \!\!\!\!\!\! 
F_{N_1,1}(\x_1,\ldots,\x_{N_1},\y_1;\x'_1,\ldots,\x'_{N_1},\y'_1;
\alpha_1,\beta_1,\alpha_2,\beta_2,C_{N_1,1},C'_{N_1,1},D_{N_1,1},D'_{N_1,1}) = \nonumber \\
& & = e^{-\frac{\alpha_1}{2} \sum_{j=1}^{N_1} (\x_j^2+{\x'_j}^2) - 
\beta_1 \sum_{1 \le j < k}^{N_1} (\x_j \cdot \x_k + \x'_j \cdot \x'_k)} \,
e^{-\frac{\alpha_2}{2} (\y_1^2+{\y'_1}^2)} \times  \\
& & \times
e^{-\frac{1}{4}C_{N_1,1} (\X_{N_1}+\X'_{N_1})^2}
e^{-\frac{1}{4}C'_{N_1,1} (\y_1+\y'_1)^2}
e^{+\frac{1}{2}D_{N_1,1} (\X_{N_1}+\X'_{N_1})(\y_1+\y'_1)}
e^{+\frac{1}{2}D'_{N_1,1} (\X_{N_1}-\X'_{N_1})(\y_1-\y'_1)}.  \nonumber
\eeqn
Its respective constants, $C_{N_1,1}$, $C'_{N_1,1}$, $D_{N_1,1}$, and $D'_{N_1,1}$,
are required as initial conditions for the vertical reduction en route 
to evaluate 
the inter-species reduced density matrix of the mixture (\ref{DENS_AB}).
Note that the dependence of $F_{N_1,1}$ on $\beta_2$ is now implicit,
representing the situation that all but the last boson of type $2$ are integrated out. 
The final result for the constants is 
\beqn\label{FUN_N1_1_COEFF}
& & C_{N_1,1} = -\gamma^2 \frac{N_2-1}{(\alpha_2-\beta_2)+{(N_2-1)\beta_2}}, \quad
 C'_{N_1,1} = -\beta_2^2 \frac{N_2-1}{(\alpha_2-\beta_2)+{(N_2-1)\beta_2}}, \nonumber \\
& & D_{N_1,1} = \gamma \frac{\alpha_2-\beta_2}{(\alpha_2-\beta_2)+{(N_2-1)\beta_2}},
\quad
D'_{N_1,1} = \gamma,
\eeqn
where the initial conditions (\ref{F_N1_N2_IN_CON}) have been used,
see appendix \ref{APP_B} for further details.

We can now perform the vertical reduction of $F_{N_1,1}$ 
where (\ref{FUN_N1_1_COEFF}) serve
as the initial conditions for the constants to be computed.
Thus we seek for a recurrence relation
\beqn\label{FUN_N_VERTICAL}
& & \int d\x_{N_1} F_{N_1,1} = \nonumber \\
& & = e^{-\frac{\alpha_1}{2} \sum_{j=1}^{N_1-1} (\x_j^2+{\x'_j}^2) - 
\beta_1 \sum_{1 \le j < k}^{N_1-1} (\x_j \cdot \x_k + \x'_j \cdot \x'_k)} \,
e^{-\frac{\alpha_2}{2} (\y_1^2+{\y'_1}^2)} \times \nonumber \\
& & \times
e^{-\frac{1}{4}C_{N_1,1} (\X_{N_1-1}+\X'_{N_1-1})^2}
e^{-\frac{1}{4}C'_{N_1,1} (\y_1+\y'_1)^2} \times \nonumber \\
& & e^{+\frac{1}{2}D_{N_1,1} (\X_{N_1-1}+\X'_{N_1-1})(\y_1+\y'_1)}
e^{+\frac{1}{2}D'_{N_1,1} (\X_{N_1-1}-\X'_{N_1-1})(\y_1-\y'_1)}\times  \nonumber \\
& & \times \int d\x_{N_1} e^{-(\alpha_1+C_{N_1,1})\x_{N_1}^2 - 
[(\beta_1+C_{N_1,1})(\X_{N_1-1}+\X'_{N_1-1}) - D_{N_1,1}(\y_1+\y'_1)]\x_{N_1}} = \nonumber \\
& & = \left(\frac{\pi}{\alpha_1+C_{N_1,1}}\right)^\frac{3}{2} 
 F_{N_1-1,1}(\x_1,\ldots,\x_{N_1-1},\y_1;\x'_1,\ldots,\x'_{N_1-1},\y'_1; \nonumber \\
& & \qquad \alpha_1,\beta_1,\alpha_2,\beta_2,C_{N_1-1,1},C'_{N_1-1,1},D_{N_1-1,1},D'_{N_1-1,1}), \ 
\eeqn
where
\beqn\label{FUN_N_VERTICAL_COEFF_1}
& & C_{N_1-1,1} = C_{N_1,1} - \frac{(\beta_1+C_{N_1,1})^2}{\alpha_1+C_{N_1,1}}, \nonumber \\
& & C'_{N_1-1,1} = C'_{N_1,1} - \frac{D^2_{N_1,1}}{\alpha_1+C_{N_1,1}}, \nonumber \\
& & D_{N_1-1,1} = D_{N_1,1} - \frac{\beta_1+C_{N_1,1}}{\alpha_1+C_{N_1,1}}D_{N_1,1}, \nonumber \\
& & D'_{N_1-1,1} = D'_{N_1,1}.
\eeqn 
The relations between the constants in the vertical reduction (\ref{FUN_N_VERTICAL_COEFF_1})
are seen to be analogous to the relations between the constants in the horizontal reduction (\ref{FUN_N_HALF_M1_2}),
see appendix \ref{APP_B} for further details.

Thus, combining and interchanging the order of
both horizontal and vertical reductions, 
the integration of $F_{N_1,N_2}$ (\ref{F_N1_N2_MATRIX_REDUCE})
ends with the auxiliary functions
\beqn\label{FUN_N_FULL_END}
& & 
F_{1,1} = e^{-\frac{\alpha_1}{2} (\x_1^2+{\x'_1}^2)}
e^{-\frac{\alpha_2}{2} (\y_1^2+{\y'_1}^2)} 
e^{-\frac{1}{4}C_{1,1} (\x_1+\x'_1)^2}
e^{-\frac{1}{4}C'_{1,1} (\y_1+\y'_1)^2} \times \nonumber \\
& & \quad \times 
e^{+\frac{1}{2}D_{1,1} (\x_1+\x'_1)(\y_1+\y'_1)}
e^{+\frac{1}{2}D'_{1,1} (\x_1-\x'_1)(\y_1-\y'_1)},  \nonumber \\
& & F_{1,0} = e^{-\frac{\alpha_1}{2} (\x_1^2+{\x'_1}^2)} e^{-\frac{1}{4}C_{1,0} (\x_1+\x'_1)^2}, \nonumber \\
& & F_{0,1} = e^{-\frac{\alpha_2}{2} (\y_1^2+{\y'_1}^2)} e^{-\frac{1}{4}C'_{0,1} (\y_1+\y'_1)^2}. \
\eeqn
Because $F_{N_1,N_2}$ with the initial conditions (\ref{F_N1_N2_IN_CON}) 
is proportional to the $N$-particle density matrix (\ref{MB_DENS}),
$F_{1,1}$ is proportional to the inter-species reduced density matrix (\ref{DENS_AB}),
and similarly $F_{1,0}$ and $F_{0,1}$ to the intra-species ones (\ref{DENS_A_B}).
The final expressions for the constants in $F_{1,1}$ are
\beqn\label{FINAL_COEFF_1}
& & C_{1,1} = 
\frac{(\alpha_1-\beta_1)C_{N_1,1}-(N_1-1)(C_{N_1,1}+\beta_1)\beta_1}{(\alpha_1-\beta_1)+(N_1-1)(C_{N_1,1}+\beta_1)},
\nonumber \\
& & C'_{1,1} = 
\frac{(\alpha_2-\beta_2)C'_{1,N_2}-(N_2-1)(C'_{1,N_2}+\beta_2)\beta_2}{(\alpha_2-\beta_2)+(N_2-1)(C'_{1,N_2}+\beta_2)},
\nonumber \\
& & D_{1,1} = \gamma \frac{(\alpha_1-\beta_1)(\alpha_2-\beta_2)}
{[(\alpha_1-\beta_1)+(N_1-1)\beta_1][(\alpha_2-\beta_2)+(N_2-1)\beta_2]-\gamma^2(N_1-1)(N_2-1)}, \nonumber \\
& & D'_{1,1} = \gamma
\eeqn
and for the constants in $F_{1,0}$ and $F_{0,1}$ are
\beqn\label{FINAL_COEFF_2}
& & C_{1,0} =
\frac{(\alpha_1-\beta_1)C_{N_1,0}-(N_1-1)(C_{N_1,0}+\beta_1)\beta_1}{(\alpha_1-\beta_1)+(N_1-1)(C_{N_1,0}+\beta_1)},
\nonumber \\
& & C'_{0,1} =  
\frac{(\alpha_2-\beta_2)C'_{0,N_2}-(N_2-1)(C'_{0,N_2}+\beta_2)\beta_2}{(\alpha_2-\beta_2)+(N_2-1)(C'_{0,N_2}+\beta_2)}. \
\eeqn
The explicit dependence of the bottommost 
constants (\ref{FINAL_COEFF_1}) and (\ref{FINAL_COEFF_2}) on
$C_{N_1,1}$ in (\ref{FUN_N1_1_COEFF}) and  
\beqn\label{FINAL_COEFF_3}
& & \qquad \qquad C'_{1,N_2} = - \gamma^2 \frac{N_1-1}{(\alpha_1-\beta_1)+(N_1-1)\beta_1}, \nonumber \\
& & C_{N_1,0} = - \gamma^2 \frac{N_2}{(\alpha_2-\beta_2)+N_2\beta_2}, \quad
 C'_{0,N_2} = - \gamma^2 \frac{N_1}{(\alpha_1-\beta_1)+N_1\beta_1} \
\eeqn
is to remind us that both horizontal and 
vertical reductions have been combined to evaluate the former,
see appendix \ref{APP_B} for further details.
All in all, the inter-species (\ref{DENS_AB}) and intra-species (\ref{DENS_A_B}) 
reduced density matrices read
\beqn\label{TWO_B_REDUCED_DENSITY}
& & 
\rho_{12}(\x,\x',\y,\y) = N_1 N_2 \left[\frac{(\alpha_1+C_{1,1})(\alpha_2+C'_{1,1})-D_{1,1}^2}{\pi^2}\right]^\frac{3}{2}
 e^{-\frac{\alpha_1}{2} (\x_1^2+{\x'_1}^2)}
e^{-\frac{\alpha_2}{2} (\y_1^2+{\y'_1}^2)} \times \nonumber \\
& & \quad \times e^{-\frac{1}{4}C_{1,1} (\x_1+\x'_1)^2}
e^{-\frac{1}{4}C'_{1,1} (\y_1+\y'_1)^2} 
e^{+\frac{1}{2}D_{1,1} (\x_1+\x'_1)(\y_1+\y'_1)}
e^{+\frac{1}{2}D'_{1,1} (\x_1-\x'_1)(\y_1-\y'_1)}  \
\eeqn
and
\beqn\label{ONE_B_REDUCED_DENSITIES}
& & \rho_1(\x,\x') = N_1 
\left(\frac{\alpha_1+C_{1,0}}{\pi}\right)^\frac{3}{2}
e^{-\frac{\alpha_1}{2} (\x_1^2+{\x'_1}^2)} e^{-\frac{1}{4}C_{1,0} (\x_1+\x'_1)^2}, \nonumber \\
& & \rho_2(\y,\y') = N_2
\left(\frac{\alpha_2+C'_{0,1}}{\pi}\right)^\frac{3}{2}
e^{-\frac{\alpha_2}{2} (\y_1^2+{\y'_1}^2)} e^{-\frac{1}{4}C'_{0,1} (\y_1+\y'_1)^2}. \
\eeqn
For completeness, we paste below their diagonal parts, i.e., the two-body
\beqn\label{TWO_B_DENSITY}
\!\!\!\!\!\!\!\!
& & \rho_{12}(\x,\y) = N_1 N_2 \left[\frac{(\alpha_1+C_{1,1})(\alpha_2+C'_{1,1})-D_{1,1}^2}{\pi^2}\right]^\frac{3}{2}
e^{-(\alpha_1+C_{1,1}) \x^2}
e^{-(\alpha_2+C'_{1,1}) \y^2}
e^{+2D_{1,1}\x \cdot \y}
\eeqn
and one-body
\beqn\label{ONE_B_DENSITIES}
& & \rho_1(\x) = N_1 
\left(\frac{\alpha_1+C_{1,0}}{\pi}\right)^\frac{3}{2}
e^{-(\alpha_1+C_{1,0})\x^2}, 
\quad 
\rho_2(\y) = N_2
\left(\frac{\alpha_2+C'_{0,1}}{\pi}\right)^\frac{3}{2}
e^{-(\alpha_2+C'_{0,1})\y^2}, \
\eeqn
densities. 
Summarizing, we have derived closed-form expressions for the (lowest-order) intra-species and inter-species 
reduced density matrices
of a generic model of a trapped mixture of interacting bosons.

\subsection{Solution of the harmonic-interaction model for trapped mixtures at the mean-field level}\label{GP_BB_GEN}

At the other end of the exact, many-body treatment of the harmonic-interaction model for trapped  
mixtures lies the Gross-Pitaevskii, mean-field solution.
In the mean-field theory the many-particle wave-function is approximated as a product state,
where all the bosons of species $1$ lie in one orbital $\phi_1(\x)$
and all the bosons of species $2$ lie in another orbital $\phi_2(\y)$.
Thus, the mean-field ansatz for the mixture is the 
product wave-function
\beq\label{MIX_WAV_GP}
 \Phi^{GP}(\x_1,\ldots,\x_{N_1},\y_1,\ldots,\y_{N_2}) = 
\prod_{j=1}^{N_1} \phi_1(\x_j) \prod_{k=1}^{N_2} \phi_2(\y_k). 
\eeq
The Gross-Pitaevskii energy functional of the mixture reads
\beqn\label{MIX_EF_GP_1}
& & E^{GP} =
N_1\Bigg[\int d\x \phi_1^\ast(\x) \left(-\frac{1}{2m_1} \frac{\partial^2}{\partial \x^2} 
+ \frac{1}{2} m_1\omega^2 \x^2 \right) \phi_1(\x) + \nonumber \\
& & \quad + \frac{\Lambda_1}{2} \int d\x d\x' |\phi_1(\x)|^2 |\phi_1(\x')|^2 (\x-\x')^2 +
\frac{\Lambda_{21}}{2}\int d\x d\y |\phi_1(\x)|^2 |\phi_2(\y)|^2 (\x-\y)^2 \Bigg] + \nonumber \\
& & \quad + N_2\Bigg[\int d\y \phi_2^\ast(\y) \left(-\frac{1}{2m_2} \frac{\partial^2}{\partial \y^2} 
+ \frac{1}{2} m_2\omega^2 \y^2 \right) \phi_2(\y) + \nonumber \\
& & \quad + \frac{\Lambda_2}{2} \int d\y d\y' |\phi_2(\y)|^2 |\phi_2(\y')|^2 (\y-\y')^2 
+ \frac{\Lambda_{12}}{2} \int d\x d\y |\phi_1(\x)|^2 |\phi_2(\y)|^2 (\x-\y)^2\Bigg], \
\eeqn 
where the mean-field interaction parameters are given by 
$\Lambda_1=\lambda_1(N_1-1)$,
$\Lambda_2=\lambda_1(N_2-1)$,
$\Lambda_{12}=\lambda_{12}N_1$, and $\Lambda_{21}=\lambda_{12}N_2$
and satisfy $N_1 \Lambda_{21} = N_2 \Lambda_{12}$.
We denote hereafter $\varepsilon^{GP}=\frac{E^{GP}}{N}$ as the 
total mean-field energy of the mixture divided by
the total number of particles $N=N_1+N_2$.
Minimizing the energy functional (\ref{MIX_EF_GP_1}) with respect to
the shapes of the orbitals $\phi_1(\x)$ and $\phi_2(\y)$, 
the two-coupled Gross-Pitaevskii equations of the mixture are derived,
\beqn\label{MIX_EQ_GP_1}
& & \left\{ -\frac{1}{2m_1} \frac{\partial^2}{\partial \x^2} + \frac{1}{2} m_1 \omega^2 \x^2 
+  \int d\x' [\Lambda_1 |\phi_1(\x')|^2 + \Lambda_{21} |\phi_2(\x')|^2] (\x-\x')^2\right\} \phi_1(\x) = \mu_1 \phi_1(\x),
\nonumber \\
& & \left\{ -\frac{1}{2m_2} \frac{\partial^2}{\partial \y^2} + \frac{1}{2} m_2 \omega^2 \y^2 
+  \int d\y' [\Lambda_2 |\phi_2(\y')|^2 + \Lambda_{12} |\phi_1(\y')|^2] (\y-\y')^2\right\} \phi_2(\y) = \mu_2 \phi_2(\y),
\nonumber \\
& & \
\eeqn
where $\mu_1$ and $\mu_2$ are the chemical potentials of the species,
see appendix \ref{APP_C}.

The solution of 
(\ref{MIX_EQ_GP_1}) follows a similar strategy as for the single-species \cite{HIM_Cohen}
and symmetric-mixture {\cite{BB_HIM_SYM} harmonic-interaction models.
Expanding the interaction terms 
in (\ref{MIX_EQ_GP_1}) we find
\beqn\label{MIX_EQ_GP_2}
& & \left\{ -\frac{1}{2m_1} \frac{\partial^2}{\partial \x^2} 
+ \frac{1}{2}m_1 \left[\omega^2 + \frac{2}{m_1}(\Lambda_1+\Lambda_{21})\right]\x^2 \right\} \phi_1(\x) =
\nonumber \\
& & \quad = 
\left\{\mu_1 - \int d\x' [\Lambda_1|\phi_1(\x')|^2 + \Lambda_{21}|\phi_2(\x')|^2]\x'^2 \right\} \phi_1(\x), \nonumber \\
& & \left\{ -\frac{1}{2m_2} \frac{\partial^2}{\partial \y^2} 
+ \frac{1}{2}m_2 \left[\omega^2 + \frac{2}{m_2}(\Lambda_2+\Lambda_{12})\right]\y^2 \right\} \phi_2(\y) =
\nonumber \\
& & \quad = 
\left\{\mu_2 - \int d\y' [\Lambda_2|\phi_2(\y')|^2 + \Lambda_{12}|\phi_1(\y')|^2]\y'^2 \right\} \phi_2(\y), \
\eeqn
where, 
since $\phi_1(\x)$ and $\phi_2(\y)$ are even functions (see below),
there are no linear in $\x$, $\y$ 
terms in (\ref{MIX_EQ_GP_2}).
A particular solution of (\ref{MIX_EQ_GP_2}) 
are the following (interaction-dressed) Gaussian functions \cite{REMARK_GAUSSIAN}
\beqn\label{MIX_GP_OR}
& & \phi_1(\x) = \left(\frac{m_1}{\pi}\sqrt{\omega^2 + \frac{2}{m_1}(\Lambda_1+\Lambda_{21})}\right)^{\frac{3}{4}}
e^{-\frac{m_1}{2}\sqrt{\omega^2 + \frac{2}{m_1}(\Lambda_1+\Lambda_{21})} \x^2} = \nonumber \\
& & \quad = \left(\frac{m_1}{\pi}\sqrt{\Omega_1^2 - \frac{2\lambda_1}{m_1}}\right)^{\frac{3}{4}}
e^{-\frac{m_1}{2}\sqrt{\Omega_1^2 - \frac{2\lambda_1}{m_1}} \x^2}, \nonumber \\
& & \phi_2(\y) = \left(\frac{m_2}{\pi}\sqrt{\omega^2 + \frac{2}{m_2}(\Lambda_2+\Lambda_{12})}\right)^{\frac{3}{4}}
e^{-\frac{m_2}{2}\sqrt{\omega^2 + \frac{2}{m_2}(\Lambda_2+\Lambda_{12})} \y^2} = \nonumber \\
& & \quad = \left(\frac{m_2}{\pi}\sqrt{\Omega_2^2 - \frac{2\lambda_2}{m_2}}\right)^{\frac{3}{4}}
e^{-\frac{m_2}{2}\sqrt{\Omega_2^2 - \frac{2\lambda_2}{m_2}} \y^2}. \
\eeqn
We can now compute
the mean-field energy per particle $\varepsilon^{GP}$
which reads
\beqn
& & \varepsilon^{GP} = 
\frac{E^{GP}}{N} = \frac{3}{2}\left[\frac{N_1}{N}\sqrt{\omega^2 + \frac{2}{m_1}(\Lambda_1+\Lambda_{21})} +
\frac{N_2}{N}\sqrt{\omega^2 + \frac{2}{m_2}(\Lambda_2+\Lambda_{12})}\right] = \nonumber \\
& &\quad
= \frac{3}{2(\Lambda_{12}+\Lambda_{21})}\left[\Lambda_{12}\sqrt{\omega^2 + \frac{2}{m_1}(\Lambda_1+\Lambda_{21})} +
\Lambda_{21}\sqrt{\omega^2 + \frac{2}{m_2}(\Lambda_2+\Lambda_{12})}\right], \
\eeqn
where $\frac{N_1}{N_2}=\frac{\Lambda_{12}}{\Lambda_{21}}$ is used.
Of course, the many-body energy (\ref{HIM_MIX_GS_E}) is always lower
than the mean-field energy because of the variational principle.

We now discuss the reduced density matrices at the mean-field level.
The Gross-Pitaevskii wave-function reads
\beqn\label{MIX_GP_WF_X_Y}
& & \Phi^{GP}(\x_1,\ldots,\x_{N_1},\y_1,\ldots,\y_{N_2}) = \nonumber \\
& & \quad = \left(\frac{m_1}{\pi}\sqrt{\omega^2 + \frac{2}{m_1}(\Lambda_1+\Lambda_{21})}\right)^{\frac{3N_1}{4}}
e^{-\frac{m_1}{2}\sqrt{\omega^2 + \frac{2}{m_1}(\Lambda_1+\Lambda_{21})} \sum_{j=1}^{N_1} \x_j^2} \times \nonumber \\
& & \quad \times 
\left(\frac{m_2}{\pi}\sqrt{\omega^2 + \frac{2}{m_2}(\Lambda_2+\Lambda_{12})}\right)^{\frac{3N_2}{4}} 
e^{-\frac{m_2}{2} \sqrt{\omega^2 + \frac{2}{m_2}(\Lambda_2+\Lambda_{12})} \sum_{k=1}^{N_2} \y_k^2}. \ 
\eeqn
From (\ref{MIX_GP_WF_X_Y}) we have
\beqn\label{MIX_GP_REDUCED_DENS_1_2}
& & \rho_1^{MF}(\x,\x') = N_1 \rho_1^{GP}(\x,\x') =
N_1 \phi_1^{GP}(\x) \left\{{\phi_1^{GP}(\x')}\right\}^\ast = \nonumber \\
& & \quad = N_1\left(\frac{m_1}{\pi}\sqrt{\omega^2 + \frac{2}{m_1}(\Lambda_1+\Lambda_{21})}\right)^{\frac{3}{2}}
e^{-\frac{m_1}{2}\sqrt{\omega^2 + \frac{2}{m_1}(\Lambda_1+\Lambda_{21})} (\x^2+{\x'}^2)} = \nonumber \\
& & \quad = N_1\left(\frac{m_1}{\pi}\sqrt{\Omega_1^2 - \frac{2\lambda_1}{m_1}}\right)^{\frac{3}{2}}
e^{-\frac{m_1}{2}\sqrt{\Omega_1^2 - \frac{2\lambda_1}{m_1}} (\x^2+{\x'}^2)}, \nonumber \\
& & \rho_2^{MF}(\y,\y') = N_2 \rho_2^{GP}(\y,\y') =
N_2 \phi_2^{GP}(\y) \left\{{\phi_2^{GP}(\y')}\right\}^\ast = \nonumber \\
& & \quad = N_2\left(\frac{m_2}{\pi}\sqrt{\omega^2 + \frac{2}{m_2}(\Lambda_2+\Lambda_{12})}\right)^{\frac{3}{2}}
e^{-\frac{m_2}{2}\sqrt{\omega^2 + \frac{2}{m_2}(\Lambda_2+\Lambda_{12})} (\y^2+{\y'}^2)} = \nonumber \\
& & \quad = N_2\left(\frac{m_2}{\pi}\sqrt{\Omega_2^2 - \frac{2\lambda_2}{m_2}}\right)^{\frac{3}{2}}
e^{-\frac{m_2}{2}\sqrt{\Omega_2^2 - \frac{2\lambda_2}{m_2}} (\y^2+{\y'}^2)} \
\eeqn
for the reduced one-body density matrices and
\beqn\label{MIX_GP_REDUCED_DENS_12}
& & \rho_{12}^{MF}(\x,\x',\y,\y') = 
N_1N_2 \phi_1^{GP}(\x) \left\{{\phi_1^{GP}(\x')}\right\}^\ast\phi_2^{GP}(\y) \left\{{\phi_2^{GP}(\y')}\right\}^\ast = \nonumber \\
& & \quad = N_1N_2\left(\frac{m_1}{\pi}\sqrt{\omega^2 + \frac{2}{m_1}(\Lambda_1+\Lambda_{21})}\right)^{\frac{3}{2}}
\left(\frac{m_2}{\pi}\sqrt{\omega^2 + \frac{2}{m_2}(\Lambda_2+\Lambda_{12})}\right)^{\frac{3}{2}} \times \nonumber \\
& & \quad \times
e^{-\frac{m_1}{2}\sqrt{\omega^2 + \frac{2}{m_1}(\Lambda_1+\Lambda_{21})} (\x^2+{\x'}^2)} 
e^{-\frac{m_2}{2}\sqrt{\omega^2 + \frac{2}{m_2}(\Lambda_2+\Lambda_{12})} (\y^2+{\y'}^2)}
= \nonumber \\
& & \quad = N_1N_2\left(\frac{m_1}{\pi}\sqrt{\Omega_1^2 - \frac{2\lambda_1}{m_1}}\right)^{\frac{3}{2}}
\left(\frac{m_2}{\pi}\sqrt{\Omega_2^2 - \frac{2\lambda_2}{m_2}}\right)^{\frac{3}{2}} \times \nonumber \\
& & \quad \times
e^{-\frac{m_1}{2}\sqrt{\Omega_1^2 - \frac{2\lambda_1}{m_1}} (\x^2+{\x'}^2)}
e^{-\frac{m_2}{2}\sqrt{\Omega_2^2 - \frac{2\lambda_2}{m_2}} (\y^2+{\y'}^2)} = \nonumber \\
& & \quad = N_1N_2 \rho_1^{GP}(\x,\x')\rho_2^{GP}(\y,\y') \
\eeqn
for the inter-species reduced density matrix.
Quite generally and as might have been expected,
for mixtures with a finite number of particles, 
the mean-field reduced density matrices (\ref{MIX_GP_REDUCED_DENS_1_2}) and (\ref{MIX_GP_REDUCED_DENS_12})  
differ from their many-body counterparts (\ref{ONE_B_REDUCED_DENSITIES}) and (\ref{TWO_B_REDUCED_DENSITY}).
The intra-species reduced density matrices are factorized to products of Gross-Pitaevskii orbitals,
and the inter-species reduced density matrix is factorized to product of Gross-Pitaevskii intra-species reduced density matrices.
This concludes our derivation of the mean-field solution
of the harmonic-interaction model for trapped mixtures.

\subsection{The infinite-particle limit}\label{INF_MB_GP_MIX}

We are now in the position to put together the above two sections,
and investigate the energy and reduced density matrices per particle of the mixture at the infinite-particle limit,
and how these quantities 
are connected in this limit with the Gross-Pitaevskii solution of the mixture.
Interestingly, for a mixture we can discuss separately two such limits,
hereafter referred to as 
the two-species infinite-particle limit and the one-species infinite-particle limit.
In the first, the numbers of particles of both species are taken to infinity whereas
in the second the number of particles of one of the species is taken to infinity and
the number of particles of the second species remains fixed and finite 
(in both limits interaction parameters are held fixed, the precise way is discussed below).
This is unlike the case of the single-species and symmetric-mixture harmonic-interaction models.
We compare and contrast the properties of the mixture in the two limits.
We start with the two-species infinite-particle limit.

\subsubsection{The two-species infinite-particle limit}\label{INF_MB_GP_MIX_II}

In the two-species infinite-particle limit,
namely for 
$N_1\to \infty$ and $N_2\to \infty$
and holding the interaction parameters
$\Lambda_1$, $\Lambda_2$, $\Lambda_{12}$, and $\Lambda_{21}$ fixed,
we find from (\ref{HIM_MIX_GS_E}) that
\beq\label{MIX_E_GP_INF}
 \lim_{N\to \infty} \frac{E}{N} = 
\frac{3}{2(\Lambda_{12}+\Lambda_{21})}\left[\Lambda_{12}\sqrt{\omega^2 + \frac{2}{m_1}(\Lambda_1+\Lambda_{21})} +
\Lambda_{21}\sqrt{\omega^2 + \frac{2}{m_2}(\Lambda_2+\Lambda_{12})}\right] =
\varepsilon^{GP},
\eeq
which establishes the connection between the exact energy per particle and mean-field (Gross-Pitaevskii) 
energy per particle in this limit for the generic mixture.
Like the literature cases of single-species bosons and the symmetric-mixture 
harmonic-interaction model,
the many-body and mean-field solutions coincide in the limit of an infinite number of particles
as far as the energy per particle in examined.
Note that the two-species limit of an infinite number of particles $N\to \infty$ implies that $N_1\to \infty$ and $N_2\to \infty$ such the
the ratio $\frac{N_1}{N_2}$ is kept constant.
This is since $\frac{N_1}{N_2}=\frac{\Lambda_{12}}{\Lambda_{21}}$.

To discuss the two-species infinite-particle limit for the reduced density matrices we first have
to evaluate the limit of relevant quantities.
Thus we find for the frequencies (\ref{MIX_FREQ}) 
\beqn\label{MIX_LIMIT_COEFFS_1}
& & \lim_{N \to \infty}\Omega_1 = \sqrt{\omega^2 + \frac{2}{m_1}(\Lambda_1+\Lambda_{21})}, \quad
\lim_{N \to \infty}\Omega_2 = \sqrt{\omega^2 + \frac{2}{m_2}(\Lambda_2+\Lambda_{21})}, \nonumber \\
& & 
\quad \lim_{N \to \infty}\Omega_{12} = \sqrt{\omega^2 + 2\left(\frac{\Lambda_{12}}{m_2}+\frac{\Lambda_{21}}{m_1}\right)}, \
\eeqn
for the parameters (\ref{WAVE_FUN_3}) of the wave-function
\beqn\label{MIX_LIMIT_COEFFS_2}
& & \lim_{N \to \infty}\alpha_1 = m_1 \sqrt{\omega^2 + \frac{2}{m_1}(\Lambda_1+\Lambda_{21})}, \quad
\lim_{N \to \infty}\alpha_2 = m_2 \sqrt{\omega^2 + \frac{2}{m_2}(\Lambda_2+\Lambda_{12})}, \nonumber \\
& & \quad \lim_{N \to \infty}\beta_1 = 0, \quad \lim_{N \to \infty}\beta_2 = 0, \quad \lim_{N \to \infty}\gamma = 0, \
\eeqn
and therefore for the constants (\ref{FINAL_COEFF_1}) and (\ref{FINAL_COEFF_2}) of the reduced density matrices 
\beqn\label{MIX_LIMIT_COEFFS_3}
& & \lim_{N \to \infty} C_{11} = 0, \quad \lim_{N \to \infty} C'_{11} = 0, \quad \lim_{N \to \infty} D_{11} = 0, \quad 
\lim_{N \to \infty} D'_{11} = 0, \nonumber \\
& & \quad \lim_{N \to \infty} C_{10} = 0, \quad \lim_{N \to \infty} C'_{01} = 0. \
\eeqn
In particular, that $\lim_{N \to \infty} D_{11} = \lim_{N \to \infty} D'_{11} = 0$
stems from $\lim_{N \to \infty}\gamma = 0$
and implies that there is no coupling at the level of the inter-species reduced density matrix, see (\ref{TWO_B_REDUCED_DENSITY}), 
between the species $1$ and $2$ 
in the two-species infinite-particle limit.
Combining the above we find
\beqn\label{1B_REDUCED_DENS_LIM}
& &  \lim_{N \to \infty} \frac{\rho_1(\x,\x')}{N_1} = 
\left(\frac{m_1}{\pi}\sqrt{\omega^2 + \frac{2}{m_1}(\Lambda_1+\Lambda_{21})}\right)^{\frac{3}{2}} 
e^{-\frac{m_1}{2}\sqrt{\omega^2 + \frac{2}{m_1}(\Lambda_1+\Lambda_{21})} (\x^2+{\x'}^2)} = \nonumber \\
& & \quad = \rho_1^{GP}(\x,\x'), \nonumber \\
& &  \lim_{N \to \infty} \frac{\rho_2(\y,\y')}{N_2} = 
\left(\frac{m_2}{\pi}\sqrt{\omega^2 + \frac{2}{m_2}(\Lambda_2+\Lambda_{12})}\right)^{\frac{3}{2}} 
e^{-\frac{m_2}{2}\sqrt{\omega^2 + \frac{2}{m_2}(\Lambda_2+\Lambda_{12})} (\y^2+{\y'}^2)} = \nonumber \\
& & \quad = \rho_2^{GP}(\y,\y')  \
\eeqn
for the reduced one-body density matrices per particle and
\beqn\label{1B_1B_REDUCED_DENS_LIM}
& & \lim_{N \to \infty} \frac{\rho_{12}(\x,\x',\y,\y')}{N_1N_2} = 
\left(\frac{m_1}{\pi}\sqrt{\omega^2 + \frac{2}{m_1}(\Lambda_1+\Lambda_{21})}\right)^{\frac{3}{2}} 
\left(\frac{m_2}{\pi}\sqrt{\omega^2 + \frac{2}{m_2}(\Lambda_2+\Lambda_{12})}\right)^{\frac{3}{2}}
\times \nonumber \\
& & \quad \times e^{-\frac{m_1}{2}\sqrt{\omega^2 + \frac{2}{m_1}(\Lambda_1+\Lambda_{21})} (\x^2+{\x'}^2)} 
e^{-\frac{m_2}{2}\sqrt{\omega^2 + \frac{2}{m_2}(\Lambda_2+\Lambda_{12})} (\y^2+{\y'}^2)} = \nonumber \\
& & \quad = \lim_{N \to \infty} \frac{\rho_1(\x,\x')}{N_1} \lim_{N \to \infty} \frac{\rho_2(\y,\y')}{N_2} = \nonumber \\
& & \quad = \rho_1^{GP}(\x,\x') \rho_2^{GP}(\y,\y') \
\eeqn
for the intra-species reduced density matrix per particle.
With this, we have established the 100\% condensation 
of each species in the generic mixture,
in the two-species infinite-particle limit.
Furthermore, the inter-species reduced density matrix per particle 
is separable in this limit and given
as a product of the intra-species reduced density matrices per particle.
Each condensate is described by the Gross-Pitaevskii quantities.
This constitutes a generalization for generic mixtures of interacting bosons, 
at least within the exactly-solvable harmonic-interaction model for trapped mixtures,
of what is known in the literature for single-species 
trapped Bose-Einstein condensates \cite{Yngvason_PRA,Lieb_PRL}.

\subsubsection{The one-species infinite-particle limit}\label{INF_MB_GP_MIX_I}

Let us discuss what happens in (and how to define) 
the limit of an infinite number of particles of one of the species,
say, species $1$. 
For $N_1 \to \infty$ the interaction parameters 
$\Lambda_1=\lambda_1(N_1-1)$ and $\Lambda_{12}=\lambda_{12}N_1$ 
are held fixed
by diminishing the interaction strengths $\lambda_1$ and $\lambda_{12}$ accordingly.
Since the number of particles of the second species $N_2$ is finite (and fixed) 
the interaction parameter $\Lambda_2=\lambda_2(N_2-1)$ 
is fixed for constant $\lambda_2$.
However, $\Lambda_{21}=\lambda_{12}N_2 \to 0$. 
Thus, we get for the energy per particle in the one-species infinite-particle limit 
\beq\label{MIX_E_GP_INF_SINGLE}
 \lim_{N_1\to \infty} \frac{E}{N} = 
\frac{3}{2} \sqrt{\omega^2 + \frac{2\Lambda_1}{m_1}} = \varepsilon^{GP}.
\eeq
When only species $1$ is taken to the infinite-particle limit it naturally becomes dominant over species $2$.
The energy per particle is that of species $1$ only,
with apparently no contribution or influence from species $2$,
and is given by the Gross-Pitaevskii energy per particle of species $1$ alone.  
What happens then with the reduced density matrices? 

To discuss the one-species infinite-particle limit for the reduced density matrices we first have
to evaluate with some care the limit of the relevant quantities.
Now we find for the frequencies (\ref{MIX_FREQ}) 
\beqn\label{MIX_LIMIT_COEFFS_1_SINGLE}
& & \lim_{N_1 \to \infty}\Omega_1 = \sqrt{\omega^2 + \frac{2\Lambda_1}{m_1}}, \quad
\lim_{N_1 \to \infty}\Omega_2 = \sqrt{\omega^2 + \frac{2}{m_2}(\Lambda_2+\lambda_2+\Lambda_{12})}, \nonumber \\
& & 
\quad \lim_{N_1 \to \infty}\Omega_{12} = \sqrt{\omega^2 + \frac{2\Lambda_{12}}{m_2}}, \
\eeqn
for the parameters (\ref{WAVE_FUN_3}) of the wave-function
\beqn\label{MIX_LIMIT_COEFFS_2_SINGLE}
& & \lim_{N_1 \to \infty}\alpha_1 = m_1 \sqrt{\omega^2 + \frac{2\Lambda_1}{m_1}}, \quad 
\lim_{N_1 \to \infty}\beta_1 = 0, \quad \lim_{N_1 \to \infty}\gamma = 0, \nonumber \\
& & 
\quad \lim_{N_1 \to \infty}\alpha_2 = m_2\left[\left(1-\frac{1}{N_2}\right)\sqrt{\omega^2 + \frac{2}{m_2}(\Lambda_2+\lambda_2+\Lambda_{12})}
+\frac{1}{N_2}\sqrt{\omega^2 + \frac{2\Lambda_{12}}{m_2}}\right] \equiv \bar\alpha_2, \nonumber \\
& & 
\quad \lim_{N_1 \to \infty}\beta_2 = 
\frac{m_2}{N_2}\left(\sqrt{\omega^2 + \frac{2\Lambda_{12}}{m_2}}-
\sqrt{\omega^2 + \frac{2}{m_2}(\Lambda_2+\lambda_2+\Lambda_{12})}\right),  \
\eeqn
and consequently 
for the constants (\ref{FINAL_COEFF_1}) and (\ref{FINAL_COEFF_2}) of the reduced density matrices 
\beqn\label{MIX_LIMIT_COEFFS_3_SINGLE}
& & \lim_{N_1 \to \infty} C_{1,1} = 0, \quad \lim_{N_1 \to \infty} C_{1,0} = 0, \quad \lim_{N_1 \to \infty} D_{1,1} = 0, \quad 
\lim_{N_1 \to \infty} D'_{1,1} = 0,  \nonumber \\
& & \lim_{N_1 \to \infty} C'_{1,1} = 
-\frac{m_2(N_2-1)}{N_2}
\frac{\left[\sqrt{\omega^2 + \frac{2}{m_2}(\Lambda_2+\lambda_2+\Lambda_{12})}-\sqrt{\omega^2 + \frac{2\Lambda_{12}}{m_2}}\right]^2}{\sqrt{\omega^2 + \frac{2}{m_2}(\Lambda_2+\lambda_2+\Lambda_{12})}+(N_2-1)\sqrt{\omega^2 + \frac{2\Lambda_{12}}{m_2}}}\nonumber \equiv \bar C'_{1,1}, \nonumber \\
& &  \lim_{N_1 \to \infty} C'_{0,1} \equiv \bar C'_{0,1} = \bar C'_{1,1}, \
\eeqn
where the last equality stems from $\lim_{N_1 \to \infty} C'_{1,N_2}=\lim_{N_1 \to \infty} C'_{0,N_2} = 0$ 
and is instrumental in what follows.
In particular, that $\lim_{N_1 \to \infty} D_{1,1} = \lim_{N_1 \to \infty} D'_{1,1} = 0$
stems from $\lim_{N_1 \to \infty}\gamma = 0$
and implies that there is no coupling at the level of the inter-species reduced density matrix, see (\ref{TWO_B_REDUCED_DENSITY}),
between the species $1$ and $2$ {\it also}
in the one-species infinite-particle limit.
Note that the corresponding quantities associated with species $2$ do not vanish in
the one-species infinite-particle limit,
compare (\ref{MIX_LIMIT_COEFFS_1_SINGLE})-(\ref{MIX_LIMIT_COEFFS_3_SINGLE}) with
(\ref{MIX_LIMIT_COEFFS_1})-(\ref{MIX_LIMIT_COEFFS_3}).
Now we can prescribe in the one-species infinite-particle limit 
the single-particle reduced density matrices per particle,
\beqn\label{1B_REDUCED_DENS_LIM_SINGLE_1_2}
& &  \lim_{N_1 \to \infty} \frac{\rho_1(\x,\x')}{N_1} = 
\left(\frac{m_1}{\pi}\sqrt{\omega^2 + \frac{2\Lambda_1}{m_1}}\right)^{\frac{3}{2}} 
e^{-\frac{m_1}{2}\sqrt{\omega^2 + \frac{2\Lambda_1}{m_1}} (\x^2+{\x'}^2)} = \rho_1^{GP}(\x,\x'), \nonumber \\
& &  \lim_{N_1\to \infty} \frac{\rho_2(\y,\y')}{N_2} = 
\left(\frac{\bar\alpha_2+\bar C'_{0,1}}{\pi}\right)^\frac{3}{2}
e^{-\frac{\bar \alpha_2}{2} (\y_1^2+{\y'_1}^2)} e^{-\frac{1}{4}\bar C'_{0,1} (\y_1+\y'_1)^2} 
\equiv \frac{\bar\rho_2(\y,\y')}{N_2} \ne \nonumber \\
& & \quad \ne \rho_2^{GP}(\y,\y'),  \ 
\eeqn
and the inter-species reduced density matrix per particle,
\beqn\label{1B_REDUCED_DENS_LIM_SINGLE_12}
& & \lim_{N_1\to \infty} \frac{\rho_{12}(\x,\x',\y,\y')}{N_1N_2} = 
\left(\frac{m_1}{\pi}\sqrt{\omega^2 + \frac{2\Lambda_1}{m_1}}\right)^{\frac{3}{2}} 
e^{-\frac{m_1}{2}\sqrt{\omega^2 + \frac{2\Lambda_1}{m_1}} (\x^2+{\x'}^2)} \times \nonumber \\
& & \quad \times \left(\frac{\bar\alpha_2+\bar C'_{1,1}}{\pi}\right)^\frac{3}{2}
e^{-\frac{\bar\alpha_2}{2} (\y_1^2+{\y'_1}^2)} e^{-\frac{1}{4}\bar C'_{1,1} (\y_1+\y'_1)^2} = \nonumber \\
& & \quad =\lim_{N_1 \to \infty} \frac{\rho_1(\x,\x')}{N_1} \lim_{N_1 \to \infty} \frac{\rho_2(\y,\y')}{N_2} = 
\rho_1^{GP}(\x,\x') \frac{\bar\rho_2(\y,\y')}{N_2} \ne \nonumber \\ 
& & \quad \ne \rho_1^{GP}(\x,\x') \rho_2^{GP}(\y,\y'). \
\eeqn
We find that 
species $1$ is described in this limit by the Gross-Pitaevskii quantity whereas species $2$, 
as might have been expected, is not,
implying that species $2$ remains correlated.
Interestingly, the inter-species reduced density matrix per particle is separable in this limit,
and precisely given by the product of the Gross-Pitaevskii quantity 
for species $1$ and the correlated quantity for species $2$, see (\ref{1B_REDUCED_DENS_LIM_SINGLE_1_2}).
This is on the account of the last equality in (\ref{MIX_LIMIT_COEFFS_3_SINGLE}).
This concludes our studies of the reduced density matrices of a generic mixture within the
harmonic-interaction model for trapped mixtures in the single-species infinite-particle limit.

\section{Concluding Remarks}\label{CON_REM}

Are the different species in the ground state of a trapped bosonic mixture $100\%$ condensed?
Are the many-body and mean-field energies per particle  
of a trapped bosonic mixture equal at the infinite-particle limit?
In the present work we answered these and more questions by treating an exactly-solvable model --
the harmonic-interaction model for trapped bosonic mixtures.

From the ground-state 
wave-function of the mixture we have computed
the lowest-order intra-species and inter-species reduced density matrices,
by generalizing 
Cohen and Lee \cite{HIM_Cohen} recurrence relations for the single-species harmonic-interaction model
to a generic mixture.
We have also obtained analytically the Gross-Pitaevskii solution for the ground state
of the mixture.
Thereafter, by taking the infinite-particle limit with respect to the two species,
we were able to show that each of the species is indeed $100\%$ condensed,
and that the many-body and Gross-Pitaevskii quantities for
the energy per particle 
and reduced density matrices per particle coincide
in this limit.
When the infinite-particle limit is taken with respect to one of the species,
only this species becomes $100\%$ condensed, 
whereas the other species remains correlated.
Interestingly, in either of the infinite-particle-limit procedures
the intra-species density matrix per particle 
becomes exactly the product of the intra-species
reduced density matrices per particle. 

It would be interesting to investigate other properties of the harmonic-interaction 
model for mixtures presented in the present work, 
e.g., quantities whose many-body and mean-field descriptions may not coincide in the infinite-particle limit 
\cite{Variance,WAVE_OVLP}.
For finite systems,
the model may prove deductive as well,
for instance to investigate properties of an impurity made of a few interacting particles 
embedded inside a larger Bose-Einstein condensate,
and to benchmark numerical tools.

\section*{Acknowledgements}

I thank Shachar Klaiman, Alexej Streltsov, and Lorenz Cederbaum for fruitful discussions. 
This research was supported by the Israel Science Foundation (Grant No. 600/15). 

\appendix

\section{Further Details of Diagonalizing the Hamiltonain of the Mixture}\label{APP_A}

Opening the braces of the particle-particle interaction terms in (\ref{HAM_MIX}) 
and collecting the diagonal contributions together we have
\beqn\label{HAM_MIX_open}
& & \hat H =
 \sum_{j=1}^{N_1} \left\{ -\frac{1}{2m_1} \frac{\partial^2}{\partial \x_j^2} +
\frac{1}{2}\left[m_1 \omega^2 + 2(N_1-1)\lambda_1 + 2N_2\lambda_{12}\right] \x_j^2 \right\} + \nonumber \\
& & \ \ \ \ + \sum_{j=1}^{N_2} \left\{ -\frac{1}{2m_2} \frac{\partial^2}{\partial \y_j^2} +
\frac{1}{2}\left[m_2 \omega^2 + 2(N_2-1)\lambda_2 + 2N_1\lambda_{12}\right] \y_j^2 \right\} - \nonumber \\
& & \ \ \ \ - 2\lambda_1 \sum_{1 \le j < k}^{N_1} \x_j \cdot \x_k 
- 2\lambda_2 \sum_{1 \le j < k}^{N_2} \y_j \cdot \y_k  
- 2\lambda_{12} \sum_{j=1}^{N_1} \sum_{k=1}^{N_2} \x_j \cdot \y_k.
\eeqn
With the Jacobi-coordinate transformation (\ref{MIX_COOR}) the 
harmonic trapping and kinetic energy remain diagonal, 
since from
\beqn\label{MIX_KIN_1}
& & \sum_{j=1}^{N_1} \x_j^2 = \sum_{k=1}^{N_1-1} \Q_k^2 
+ \left(\frac{\sqrt{N_2}m_2}{M}\Q_{N-1}+\sqrt{N_1}\Q_N\right)^2 = \nonumber \\
& & \qquad = \sum_{k=1}^{N_1-1} \Q_k^2 
+ \frac{N_2m_2^2}{M^2} \Q_{N-1}^2 + N_1 \Q_N^2 
+ \frac{2\sqrt{N_1N_2}m_2}{M} \Q_{N-1} \Q_N, 
\nonumber \\
& & \sum_{j=1}^{N_2} \y_j^2 = \sum_{k=N_1}^{N-2} \Q_k^2 
+ \left(-\frac{\sqrt{N_1}m_1}{M}\Q_{N-1} + \sqrt{N_2}\Q_N\right)^2 = \nonumber \\
& & \qquad = \sum_{k=N_1}^{N-2} \Q_k^2 
+ \frac{N_1m_1^2}{M^2}\Q_{N-1}^2 + N_2 \Q_N^2 
- \frac{2\sqrt{N_1N_2}m_1}{M} \Q_{N-1} \Q_N \
\eeqn
and
\beqn\label{MIX_KIN_1p5}
& & \sum_{j=1}^{N_1} \frac{\partial^2}{\partial \x_j^2} = \sum_{k=1}^{N_1-1} \frac{\partial^2}{\partial \Q_k^2} +  \left(\sqrt{N_2} \frac{\partial}{\partial \Q_{N-1}} + \frac{\sqrt{N_1}m_1}{M}\frac{\partial}{\partial \Q_N}\right)^2 = \nonumber \\
& & \qquad = 
\sum_{k=1}^{N_1-1} \frac{\partial^2}{\partial \Q_k^2} + N_2 \frac{\partial^2}{\partial \Q_{N-1}^2} + 
\frac{N_1m_1^2}{M^2}\frac{\partial^2}{\partial \Q_N^2} + 
\frac{2\sqrt{N_1N_2}m_1}{M} \frac{\partial}{\partial \Q_{N-1}} \frac{\partial}{\partial \Q_N}, \nonumber \\
& & \sum_{j=1}^{N_2} \frac{\partial^2}{\partial \y_j^2} = 
\sum_{k=N_1}^{N-2} \frac{\partial^2}{\partial \Q_k^2} + 
\left(-\sqrt{N_1}\frac{\partial}{\partial \Q_{N-1}} + \frac{\sqrt{N_2}m_2}{M} \frac{\partial}{\partial \Q_N} \right)^2 = \nonumber \\
& & \qquad = \sum_{k=N_1}^{N-2} \frac{\partial^2}{\partial \Q_k^2} + N_1 \frac{\partial^2}{\partial \Q_{N-1}^2} + 
\frac{N_2m_2^2}{M^2} \frac{\partial^2}{\partial \Q_N^2} - 
\frac{2\sqrt{N_1N_2}m_2}{M} \frac{\partial}{\partial \Q_{N-1}} \frac{\partial}{\partial \Q_N} \
\eeqn
we have
\beqn\label{MIX_KIN_2}
& & m_1\sum_{j=1}^{N_1} \x_j^2 + m_2\sum_{j=1}^{N_2} \y_j^2 =
m_1\sum_{k=1}^{N_1-1} \Q_k^2 + m_2\sum_{k=N_1}^{N-2} \Q_k^2 + 
M_{12}\Q_{N-1}^2 + M\Q_N^2, \nonumber \\
& & 
\frac{1}{m_1}\sum_{j=1}^{N_1} \frac{\partial^2}{\partial \x_j^2} + 
\frac{1}{m_2}\sum_{j=1}^{N_2} \frac{\partial^2}{\partial \y_j^2} =
\sum_{k=1}^{N_1-1}\frac{1}{m_1}\frac{\partial^2}{\partial \Q_k^2} +
\sum_{k=N_1}^{N-2}\frac{1}{m_2}\frac{\partial^2}{\partial \Q_k^2} +
\frac{1}{M_{12}}\frac{\partial^2}{\partial \Q_{N-1}^2} +
\frac{1}{M}\frac{\partial^2}{\partial \Q_N^2}, \nonumber \\
& & \qquad M_{12} = \frac{1}{\frac{N_1}{m_2}+\frac{N_2}{m_1}} = \frac{m_1m_2}{M}, \qquad M=N_1m_1+N_2m_2. \
\eeqn
Finally, 
using the quadratic relations
\beqn\label{MIX_POT_1}
& & \sum_{k=1}^{N_1-1} \Q_k^2 = \left(1-\frac{1}{N_1}\right) \sum_{j=1}^{N_1} \x_j^2 
-\frac{2}{N_1} \sum_{1 \le j < k}^{N_1} \x_j \cdot \x_k, \nonumber \\
& & \sum_{k=N_1}^{N-2} \Q_k^2 = \left(1-\frac{1}{N_2}\right) \sum_{j=1}^{N_2} \y_j^2 
-\frac{2}{N_2} \sum_{1 \le j < k}^{N_2} \y_j \cdot \y_k, \nonumber \\
& & \Q^2_{N-1} = \frac{N_2}{NN_1} \sum_{j=1}^{N_1} \x_j^2 + \frac{N_1}{NN_2} \sum_{j=1}^{N_2} \y_j^2 \nonumber \\
& & \qquad + 2\left[\frac{N_2}{NN_1} \sum_{1 \le j < k}^{N_1} \x_j \cdot \x_k 
+ \frac{N_1}{NN_2} \sum_{1 \le j < k}^{N_2} \y_j \cdot \y_k 
- \frac{1}{N} \sum_{j=1}^{N_1} \sum_{k=1}^{N_2} \x_j \cdot \y_k\right], \nonumber \\
& & \Q^2_N = \frac{m_1^2}{M^2} \sum_{j=1}^{N_1} \x_j^2 + \frac{m_2^2}{M^2} \sum_{j=1}^{N_2} \y_j^2 \nonumber \\
& & \qquad + 2\left[\frac{m_1^2}{M^2} \sum_{1 \le j < k}^{N_1} \x_j \cdot \x_k 
+ \frac{m_2^2}{M^2} \sum_{1 \le j < k}^{N_2} \y_j \cdot \y_k 
+ \frac{m_1m_2}{M^2} \sum_{j=1}^{N_1} \sum_{k=1}^{N_2} \x_j \cdot \y_k\right], \
\eeqn
the coupling terms in the Hamiltonian (\ref{HAM_MIX_open}) are diagonalized too,
and the ground-state wave-function transforms from the Jacobi-coordinate
to the lab-frame representation,
see (\ref{WAVE_FUN_1}) and (\ref{WAVE_FUN_2}).

\section{Further Details of the Solution of the Coupled Recurrence Relations}\label{APP_B}

Within the horizontal reduction of $F_{N_1,N_2}$, see (\ref{F_N1_N2_MATRIX_REDUCE})-(\ref{FUN_N_HALF_M1_2}),
the recurrence relations between the corresponding constants are 
\beqn\label{FUN_N_HALF_M1_3}
& & C'_{N_1,j-1} = C'_{N_1,j} - \frac{(\beta_2+C'_{N_1,j})^2}{\alpha_2+C'_{N_1,j}}, \nonumber \\
& & D_{N_1,j-1} = D_{N_1,j} - \frac{\beta_2+C'_{N_1,j}}{\alpha_2+C'_{N_1,j}}D_{N_1,j} 
= \frac{\alpha_2-\beta_2}{\alpha_2+C'_{N_1,j}}D_{N_1,j} , \nonumber \\
& & C_{N_1,j-1} = C_{N_1,j} - \frac{D^2_{N_1,j}}{\alpha_2+C'_{N_1,j}} = 
C_{N_1,j} - \frac{D_{N_1,j}D_{N_1,j-1}}{\alpha_2-\beta_2}, \nonumber \\
& & D'_{N_1,j-1} = D'_{N_1,j}. \
\eeqn
The recurrence 
relations (\ref{FUN_N_HALF_M1_3}) 
consist of one non-linear relation (for $C'_{N_1,j}$)
and the linear relations for $D_{N_1,j}$, $C_{N_1,j}$, and $D'_{N_1,j}$.
We see from (\ref{FUN_N_HALF_M1_3}) that first the
relation for $C'_{N_1,j}$ needs to be solved,
then for the $D_{N_1,j}$, and then for the $C_{N_1,j}$.
The linear relation for $D'_{N_1,j}$ is trivial and does not depend on the other constants.

The recurrence relation for $C'_{N_1,j}$ 
has exactly the same structure as the recurrence relation 
emerging in the integration of the single-species harmonic-interaction model \cite{HIM_Cohen}.
Making use of this observation and substituting the result into $D_{N_1,j}$ and then together into $C_{N_1,j}$,
relations (\ref{FUN_N_HALF_M1_3}) are solved.
The final result reads
\beqn\label{FUN_N_HALF_M1_4}
& &  C'_{N_1,j} = -\alpha_2 + \frac{(\alpha_2-\beta_2)(1+j\eta_2)}{1+(j+1)\eta_2}, 
\qquad \eta_2=-\frac{\beta_2}{\alpha_2+N_2\beta_2}\nonumber \\
& & D_{N_1,j} = \gamma \frac{1+(N_2+1)\eta_2}{1+(j+1)\eta_2}, \nonumber \\
& & C_{N_1,j} = - \frac{1}{\alpha_2-\beta_2} \sum_{k=N_2}^{j+1} D_{N_1,k} D_{N_1,k-1} = 
-\frac{\gamma^2 [1+(N_2+1)\eta_2]}{\alpha_2-\beta_2} \frac{(N_2-j)}{1+(j+1)\eta_2}, \nonumber \\
& & D'_{N_1,j}=\gamma, \
\eeqn
where the initial conditions for $C'_{N_1,N_2}$, $D_{N_1,N_2}$, $C_{N_1,N_2}$, and $D'_{N_1,N_2}$ in (\ref{F_N1_N2_IN_CON})
have been used.
From (\ref{FUN_N_HALF_M1_4}) we obtain the constants 
$C'_{N_1,1}$, $D_{N_1,1}$, $C_{N_1,1}$, and $D'_{N_1,1}$ in (\ref{FUN_N1_1_COEFF}) 
entering the auxiliary function $F_{N_1,1}$,
and
$C_{N_1,0}$ in (\ref{FINAL_COEFF_3}) entering the auxiliary function $F_{N_1,0}$.

Proceeding now to the vertical reduction of $F_{N_1,1}$ to $F_{1,1}$,
 see (\ref{F_N1_N2_MATRIX_REDUCE}), (\ref{FUN_N_VERTICAL}), and (\ref{FUN_N_VERTICAL_COEFF_1}),
we write
\beqn\label{FUN_N_VERTICAL_COEFF_2}
& & C_{j-1,1} = C_{j,1} - \frac{(\beta_1+C_{j,1})^2}{\alpha_1+C_{j,1}}, \nonumber \\
& & D_{j-1,1} = D_{j,1} - \frac{\beta_1+C_{j,1}}{\alpha_1+C_{j,1}}D_{j,1}, \nonumber \\
& & C'_{j-1,1} = C'_{j,1} - \frac{D^2_{j,1}}{\alpha_1+C_{j,1}}, \nonumber \\
& & D'_{j-1,1} = D'_{j,1}.
\eeqn
The vertical recursion relations (\ref{FUN_N_VERTICAL_COEFF_2}) 
consist again one non-linear relation (for $C_{j,1}$),
and the linear relations for $C'_{j,1}$, $D_{j,1}$, and $D'_{j,1}$.
The linear relation for $D'_{j,1}$ is again trivial.
Note the interchange of roles of the $C$ and $C'$ constants when moving from horizontal to vertical reductions.
We can hence interchange the order of horizontal and vertical reductions
in order to prescribe $C'_{1,1}$ based on the solution for $C_{1,1}$.
Then, $C'_{1,N_2}$ in (\ref{FINAL_COEFF_3}) obtained 
from the vertical reduction is used as an initial condition for the recurrence relation.
Combining all the above we find for the constants of the auxiliary function $F_{1,1}$
\beqn\label{FUN_N_VERTICAL_COEFF_3}
& & C_{1,1} = -\alpha_1 + \frac{(\alpha_1-\beta_1)(1+\eta_1)}{1+2\eta_1}, \qquad
\eta_1=\frac{(\alpha_1-\beta_1) - (\alpha_1+C_{N_1,1})}{(N_1+1)(\alpha_1+C_{N_1,1})-N_1(\alpha_1-\beta_1)}, \nonumber \\
& & D_{1,1} = D_{N_1,1} \frac{1+(N_1+1)\eta_1}{1+2\eta_1}, \nonumber \\
& & C'_{1,1} = -\alpha_2 + \frac{(\alpha_2-\beta_2)(1+\eta'_1)}{1+2\eta'_1}, \qquad
\eta'_1=\frac{(\alpha_2-\beta_2) - (\alpha_2+C'_{1,N_2})}{(N_2+1)(\alpha_2+C'_{1,N_2})-N_2(\alpha_2-\beta_2)}, \nonumber \\
& & D'_{1,1} = D'_{N_1,1}.
\eeqn
Upon substitution we arrive at the final result (\ref{FINAL_COEFF_1}).

Last are the constants of $F_{1,0}$ and $F_{0,1}$.
From the recurrence relation
\beq\label{FUN_N_VERTICAL_COEFF_4}
C_{j-1,0} = C_{j,0} - \frac{(\beta_1+C_{j,0})^2}{\alpha_1+C_{j,0}}
\eeq
we find
\beqn\label{FUN_N_VERTICAL_COEFF_5}
& & C_{1,0} = -\alpha_1 + \frac{(\alpha_1-\beta_1)(1+\eta_0)}{1+2\eta_0}, \qquad
\eta_0=\frac{(\alpha_1-\beta_1) - (\alpha_1+C_{N_1,0})}{(N_1+1)(\alpha_1+C_{N_1,0})-N_1(\alpha_1-\beta_1)}, \nonumber \\
& & C'_{0,1} = -\alpha_2 + \frac{(\alpha_2-\beta_2)(1+\eta'_0)}{1+2\eta'_0}, \qquad
\eta'_0=\frac{(\alpha_2-\beta_2) - (\alpha_2+C'_{0,N_2})}{(N_2+1)(\alpha_2+C'_{0,N_2})-N_2(\alpha_2-\beta_2)}, \
\eeqn
where the initial conditions $C_{N_1,0}$ and $C'_{0,N_2}$ are given in (\ref{FINAL_COEFF_3}),
and where interchanging the order of the vertical and horizontal reductions
allows one to obtain $C'_{0,1}$ analogously to $C_{1,0}$.
Substituting all quantities we arrive at the final expressions for the constants (\ref{FINAL_COEFF_2}).

\section{Further Details of the Mean-Field Solution of the Mixture}\label{APP_C}
 
Given the orbitals $\phi_1(\x)$ and $\phi_2(\y)$ (\ref{MIX_GP_OR}), 
we can now evaluate the integrals\break\hfill $\int d\x' |\phi_1(\x')|^2 \x'^2 = 
\frac{3}{2\sqrt{\omega^2 + \frac{2}{m_1}(\Lambda_1+\Lambda_{21})}}$ 
and $\int d\y' |\phi_2(\y')|^2 \y'^2 = \frac{3}{2\sqrt{\omega^2 + \frac{2}{m_1}(\Lambda_2+\Lambda_{12})}}$
in (\ref{MIX_EQ_GP_2}) and determine the chemical potentials
\beqn\label{MIX_GP_MU}
& & \!\!\!\!\!\! \mu_1 = \frac{3}{2}\sqrt{\omega^2 + \frac{2}{m_1}(\Lambda_1+\Lambda_{21})} 
+ \frac{3}{2}\left(\frac{\Lambda_1}{\sqrt{\omega^2 + \frac{2}{m_1}(\Lambda_1+\Lambda_{21})}} 
+ \frac{\Lambda_{21}}{\sqrt{\omega^2 + \frac{2}{m_2}(\Lambda_2+\Lambda_{12})}}\right), \nonumber \\
& & \!\!\!\!\!\! \mu_2 = \frac{3}{2}\sqrt{\omega^2 + \frac{2}{m_2}(\Lambda_2+\Lambda_{12})}
+ \frac{3}{2}\left(\frac{\Lambda_2}{\sqrt{\omega^2 + \frac{2}{m_2}(\Lambda_2+\Lambda_{12})}} 
+ \frac{\Lambda_{12}}{\sqrt{\omega^2 + \frac{2}{m_1}(\Lambda_1+\Lambda_{21})}}\right). \
\eeqn 
With this, the mean-field energy is given by
\beqn\label{MIX_E_GP}
& & E^{GP} = N_1\mu_1 + N_2\mu_2 - \nonumber \\
& & \quad - \frac{N_1}{2}\left[\Lambda_1\int d\x d\x' |\phi_1(\x)|^2 |\phi_1(\x')|^2 (\x-\x')^2 
+ \Lambda_{21}\int d\x d\y |\phi_1(\x)|^2 |\phi_2(\y)|^2 (\x-\y)^2\right] - \nonumber \\
& & \quad - \frac{N_2}{2}\left[\Lambda_2\int d\y d\y' |\phi_2(\y)|^2 |\phi_2(\y')|^2 (\y-\y')^2 
+ \Lambda_{12}\int d\x d\y |\phi_1(\x)|^2 |\phi_2(\y)|^2 (\x-\y)^2\right] = \nonumber \\
& & \quad = 
\frac{3}{2}\left[N_1\sqrt{\omega^2 + \frac{2}{m_1}(\Lambda_1+\Lambda_{21})} +
N_2\sqrt{\omega^2 + \frac{2}{m_2}(\Lambda_2+\Lambda_{12})}\right] = \nonumber \\
& & \quad = \frac{3}{2}\left[N_1\sqrt{\Omega_1^2 - \frac{2\lambda_1}{m_1}} + N_2\sqrt{\Omega_2^2 - \frac{2\lambda_2}{m_2}}\right],
\eeqn
where 
$\int d\x d\x' |\phi_1(\x)|^2 |\phi_1(\x')|^2 (\x-\x')^2 = \frac{3}{\sqrt{\omega^2 + \frac{2}{m_1}(\Lambda_1+\Lambda_{21})}}$,
$\int d\y d\y' |\phi_2(\y)|^2 |\phi_2(\y')|^2 (\y-\y')^2 = \frac{3}{\sqrt{\omega^2 + \frac{2}{m_2}(\Lambda_2+\Lambda_{12})}}$,
and
$\int d\x d\y |\phi_1(\x)|^2 |\phi_2(\y)|^2 (\x-\y)^2 = 
\frac{3}{2}\left[\frac{1}{\sqrt{\omega^2 + \frac{2}{m_1}(\Lambda_1+\Lambda_{21})}} +
\frac{1}{\sqrt{\omega^2 + \frac{2}{m_2}(\Lambda_2+\Lambda_{12})}}\right]$
are used. 




\begin{thebibliography}{99}

\bibitem{B1} C. J. Myatt, E. A. Burt, R. W. Ghrist, E. A. Cornell, and C. E. Wieman,
                   {\rm Production of Two Overlapping Bose-Einstein Condensates by Sympathetic Cooling},
                   Phys. Rev. Lett. {\bf 78}, 586 (1997).

\bibitem{B2} D. M. Stamper-Kurn, M. R. Andrews, A. P. Chikkatur, S. Inouye, 
                   H.-J. Miesner, J. Stenger, and W. Ketterle,
                   {\rm Optical Confinement of a Bose-Einstein Condensate},
                   Phys. Rev. Lett. {\bf 80}, 2027 (1998).

\bibitem{B3} T.-L. Ho and V. B. Shenoy,
                    {\rm Binary Mixtures of Bose Condensates of Alkali Atoms},
                    Phys. Rev. Lett. {\bf 77}, 3276 (1996).

\bibitem{B4} B. D. Esry, C. H. Greene, J. P. Burke, Jr., and J. L. Bohn, 
                    {\rm Hartree-Fock Theory for Double Condensates},
                    Phys. Rev. Lett. {\bf 78}, 3594 (1997).

\bibitem{B5} H. Pu and N. P. Bigelow, 
                    {\rm Properties of Two-Species Bose Condensates},
                    Phys. Rev. Lett. {\bf 80}, 1130 (1998).

\bibitem{B6} E. Timmermans, 
                    {\rm Phase Separation of Bose-Einstein Condensates},
                    Phys. Rev. Lett. {\bf 81}, 5718 (1998).

\bibitem{B7} E. Altman, W. Hofstetter, E. Demler, and M. D. Lukin,
                    {\rm Phase diagram of two-component bosons on an optical lattice},
                    New J. Phys. {\bf 5}, 113 (2003).

\bibitem{B8} A. B. Kuklov and B. V. Svistunov,
                   {\rm Counterflow Superfluidity of Two-Species Ultracold 
                    Atoms in a Commensurate Optical Lattice},
                    Phys. Rev. Lett. {\bf 90}, 100401 (2003).

\bibitem{B9} A. Eckardt, C. Weiss, and M. Holthaus,
                     {\rm Ground-state energy and depletions for a dilute binary Bose gas},
                     Phys. Rev. A {\bf 70}, 043615 (2004).

\bibitem{B10} O. E. Alon, A. I. Streltsov, and L. S. Cederbaum,
                     {\rm Demixing of Bosonic Mixtures in Optical Lattices from Macroscopic to Microscopic},
                     Phys. Rev. Lett. {\bf 97}, 230403 (2006).

\bibitem{B11} O. E. Alon, A. I. Streltsov, and L. S. Cederbaum,
                     {\rm Multiconfigurational time-dependent Hartree method for 
                     mixtures consisting of two types of identical particles},
                     Phys. Rev. A. {\bf 76}, 062501 (2007).

\bibitem{B12} A. R. Sakhel, J. L. DuBois, and H. R. Glyde,
                     {\rm Condensate depletion in two-species Bose gases: A variational quantum Monte Carlo study},
                     Phys. Rev. A {\bf 77}, 043627 (2008).

\bibitem{B13} B. Ole\'s and K. Sacha,
                     {\rm $N$-conserving Bogoliubov vacuum of a two-component 
                     Bose-Einstein condensate: density fluctuations close to a phase-separation condition},
                     J. Phys. A {\bf 41}, 145005 (2008).

\bibitem{B14} Y. Hao and S. Chen,
                    {\rm Density-functional theory of two-component Bose gases in one-dimensional harmonic traps},
                     Phys. Rev. A {\bf 80}, 043608 (2009).

\bibitem{B15} M. D. Girardeau,
                     {\rm Pairing, Off-Diagonal Long-Range Order, and Quantum Phase Transition in Strongly 
                     Attracting Ultracold Bose Gas Mixtures in Tight Waveguides},
                     Phys. Rev. Lett. {\bf 102}, 245303 (2009).

\bibitem{B16} J. Smyrnakis, S. Bargi, G. M. Kavoulakis, M. Magiropoulos, K. K\"arkk\"ainen, and S. M. Reimann,
                     {\rm Mixtures of Bose Gases Confined in a Ring Potential},
                     Phys. Rev. Lett. {\bf 103}, 100404 (2009).

\bibitem{B17} M. D. Girardeau and G. E. Astrakharchik, 
                     {\rm Ground state of a mixture of two bosonic 
                     Calogero-Sutherland gases with strong odd-wave interspecies attraction},
                     Phys. Rev. A {\bf 81}, 043601 (2010).

\bibitem{B18} S. Gautam and D. Angom,
                     {\rm Ground state geometry of binary condensates in axissymmetric traps}, 
                     J. Phys. B. {\bf 43}, 095302 (2010). 

\bibitem{B19} S. Kr\"onke, L. Cao, O. Vendrell, and P. Schmelcher,
                     {\rm Non-equilibrium quantum dynamics of ultra-cold atomic mixtures: 
                     the multi-layer multi-configuration time-dependent Hartree method for bosons},
                     New J. Phys. {\bf 15}, 063018 (2013).
                    
\bibitem{B20} M. A. Garcia-March and T. Busch,
                     {\rm Quantum gas mixtures in different correlation regimes},
                     Phys. Rev. A {\bf 87}, 063633 (2013).

\bibitem{B21} K. Anoshkin, Z. Wu, and E. Zaremba,
                     {\rm Persistent currents in a bosonic mixture in the ring geometry},
                     Phys. Rev. A {\bf 88}, 013609 (2013).  

\bibitem{B22} L. Cao, S. Kr\"onke, O. Vendrell, and P. Schmelcher,
                    {\rm The multi-layer multi-configuration time-dependent Hartree method for bosons: 
                    Theory, implementation, and applications},
                    J. Chem. Phys. {\bf 139}, 134103 (2013).

\bibitem{B23} L. A. Pe\~{n}a Ardila and S. Giorgini,
                     {\rm Impurity in a Bose-Einstein condensate: 
                     Study of the attractive and repulsive branch using quantum Monte Carlo methods},
                     Phys. Rev. A {\bf 92}, 033612 (2015).

\bibitem{B24} S. Kr\"onke, J. Kn\"orzer, and P. Schmelcher, 
                    {\rm Correlated quantum dynamics of a single atom collisionally coupled to an ultracold finite bosonic ensemble},  
                    New J. Phys. {\bf 17}, 053001 (2015).

\bibitem{B25} D.  S. Petrov,
                     {\rm Quantum Mechanical Stabilization of a Collapsing Bose-Bose Mixture},
                     Phys. Rev. Lett. {\bf 115}, 155302 (2015).

\bibitem{B26} I. Anapolitanos, M. Hott, and D. Hundertmark,
                     {\rm Derivation of the Hartree equation for compound Bose gases in the mean field limit},
                     arXiv:1702.00827v1 [math-ph].

\bibitem{Yngvason_PRA} E. H. Lieb, R. Seiringer, and J. Yngvason,
                                  {\rm Bosons in a trap: A rigorous derivation of the Gross-Pitaevskii energy functional},
                                   Phys. Rev. A {\bf 61}, 043602 (2000).

\bibitem{Lieb_PRL} E. H. Lieb and R. Seiringer,
                           {\rm Proof of Bose-Einstein Condensation for Dilute Trapped Gases},
                           Phys. Rev. Lett. {\bf 88}, 170409 (2002).

 
\bibitem{HIM_Cohen} L. Cohen and C. Lee,
                              {\rm Exact reduced density matrices for a model problem}, 
                              J. Math. Phys. {\bf 26}, 3105 (1985).

\bibitem{HIM_Yan} J. Yan, 
                           {\rm Harmonic Interaction Model and Its Applications in Bose-Einstein Condensation},
                           J. Stat. Phys. {\bf 113}, 623 (2003).

\bibitem{HIM_Po1} M. Gajda, 
                           {\rm Criterion for Bose-Einstein condensation in a harmonic 
                           trap in the case with attractive interactions}, 
                           Phys. Rev. A {\bf 73}, 023603 (2006).

\bibitem{HIM_Benchmarks} A. U. J. Lode, K. Sakmann, O. E. Alon, L. S. Cederbaum, and A. I. Streltsov, 
                                      {\rm Numerically exact quantum dynamics of bosons with 
                                      time-dependent interactions of harmonic type},
                                      Phys. Rev. A {\bf 86}, 063606 (2012).


\bibitem{HIM_Po0} M. A. Za\l{}uska-Kotur, M. Gajda, A. Or\l{}owski, and J. Mostowski,
                           {\rm Soluble model of many interacting quantum particles in a trap},
                           Phys. Rev. A {\bf 61}, 033613 (2000).

\bibitem{Schilling_FER_BOS} C. Schilling,
                                         {\rm Natural orbitals and occupation numbers for harmonium: Fermions versus bosons},
                                         Phys. Rev. A {\bf 88}, 042105 (2013).

\bibitem{Ental_FER_BOS}  C. L. Benavides-Riveros, I. V. Toranzo, and J. S. Dehesa,
                                      {\rm Entanglement in N-harmonium: bosons and fermions},
                                      J. Phys. B {\bf 47}, 195503 (2014). 


\bibitem{Schilling_FER_HIM_WORK} C. Schilling and R. Schilling,
                                                  {\rm Number-parity effect for confined fermions in one dimension},
                                                  Phys. Rev. {\bf 93}, 021601(R) (2016).

\bibitem{Axel_MCTDHF_HIM} E. Fasshauer and A. U. J. Lode,
                                         {\rm Multiconfigurational time-dependent Hartree method for fermions: 
                                         Implementation, exactness, and few-fermion tunneling to open space},
                                         Phys. Rev. A {\bf 93}, 033635 (2016).

\bibitem{HIM_MIX_JPA1} R. L. Hall,
                                     {\rm Some exact solutions to the translation-invariant N-body problem},
                                     J. Phys. A {\bf 11}, 1227 (1978). 

\bibitem{HIM_MIX_JPA2} R. L. Hall,
                                     {\rm Exact solutions of Schr\"odinger's equation for translation-invariant harmonic matter},
                                      J. Phys. A {\bf 11}, 1235 (1978). 

\bibitem{HIM_MIX_PRE} M. S. Osadchii and V. V. Muraktanov,
                                  {\rm The System of Harmonically Interacting Particles: 
                                   An Exact Solution of the Quantum-Mechanical Problem},
                                   Int. J. Quant. Chem. {\bf 39}, 173 (1991).

\bibitem{BB_HIM_SYM} S. Klaiman, A. I. Streltsov, and O. E. Alon,
                                 {\rm Solvable model of a trapped mixture of Bose-Einstein condensates},
                                 Chem. Phys. {\bf 482}, 362 (2017). 

\bibitem{REMARK_INTERACTIONS} There are eight different combinations for which the particle-particle interactions 
$\lambda_1$, $\lambda_2$, and $\lambda_{12}$ are
either positive (corresponds to attraction) or negative (corresponds to repulsion).
Note that the trap potential enables four 
more combinations 
compared to the mixture in free space \cite{HIM_MIX_PRE},
since it allows for repulsion between the two species
while the system is still bound.

\bibitem{REMARK_DEGEN} In some specific cases 
frequencies can become degenerate
and their multiplicity changes.
For instance,
when the masses and interactions are inter-connected by the relations
$$
\frac{\lambda_1}{\lambda_{12}}=\frac{\lambda_{12}}{\lambda_2}=\frac{m_1}{m_2} \quad \Longleftrightarrow \quad
\sqrt{\frac{\lambda_1}{\lambda_2}}=\frac{m_1}{m_2},\, \lambda_{12}=\sqrt{\lambda_1\lambda_2},  
$$
one finds that $\Omega_1=\Omega_2=\Omega_{12}$,
i.e., all relative coordinates have the same frequency.
This degeneracy of the frequencies
holds independently of the numbers of bosons $N_1$ and $N_2$.
The ground-state energy reads
$E = \frac{3}{2} \left[ (N-1) \sqrt{\omega^2 + \frac{2\lambda_{12}}{M_{12}}} + \omega \right]$ 
and expressed as a function of the inter-species interaction only.
Another example is when
the numbers of particles 
and interactions are inter-connected by the relations
$$
N_1\lambda_1+N_2\lambda_{12}=0, \, N_2\lambda_2+N_1\lambda_{12}=0
 \quad \Longleftrightarrow \quad
\lambda_1 = -\frac{N_2}{N_1} \lambda_{12}, \,
\lambda_2 = -\frac{N_1}{N_2} \lambda_{12},
$$
one gets that $\Omega_1=\Omega_2=\omega$,
i.e., the intra-species relative coordinates have the same frequency as the center-of-mass coordinate.
This degeneracy of the frequencies 
holds independently of the masses of the bosons $m_1$ and $m_2$.
The energy now reads
$E = \frac{3}{2} \left[ (N-1) \omega + \sqrt{\omega^2 + \frac{2\lambda_{12}}{M_{12}}}\right]$
and again expressed as a function of the inter-species interaction only.

\bibitem{REMARK_NO_TRAP} In the limit of vanishing trapping frequency, $\omega=0$, 
                            the relative-motion part of $\Psi$, 
                            which is built from all relative Jacobi coordinates of the mixture $\Q_1,\ldots,\Q_{N-1}$, 
                            and the energy $E$ 
                            boil down to those of the mixture in free space \cite{HIM_MIX_PRE}.

\bibitem{REMARK_VECTORS} The relations
\beqn
& & (\X_{N_1} + \X'_{N_1})^2 = (\X_{N_1-1} + \X'_{N_1-1})^2 + 
2 (\X_{N_1-1} + \X'_{N_1-1}) (\x_{N_1} + \x'_{N_1}) + (\x_{N_1} + \x'_{N_1})^2, \nonumber \\
& & (\Y_{N_2} + \Y'_{N_2})^2 = (\Y_{N_2-1} + \Y'_{N_2-1})^2 + 
2 (\Y_{N_2-1} + \Y'_{N_2-1}) (\y_{N_2} + \y'_{N_2}) + (\y_{N_2} + \y'_{N_2})^2, \nonumber \\
& & (\X_{N_1} \pm \X'_{N_1}) (\Y_{N_2} \pm \Y'_{N_2}) =  
(\X_{N_1-1} \pm \X'_{N_1-1}) (\Y_{N_2} \pm \Y'_{N_2}) + (\x_{N_1} \pm \x'_{N_1}) (\Y_{N_2} \pm \Y'_{N_2}) = \nonumber \\
& & \quad =
(\X_{N_1} \pm \X'_{N_1}) (\Y_{N_2-1} \pm \Y'_{N_2-1}) + (\X_{N_1} \pm \X'_{N_1}) (\y_{N_2} \pm \y'_{N_2}) \nonumber
\eeqn
between the vectors introduced in (\ref{FUN_N_HALF_2}) and the 
variables of the degrees of freedom 
to be integrated $\x_{N_1}$, $\x'_{N_1}$, $\y_{N_2}$ and $\y'_{N_2}$ hold.

\bibitem{REMARK_GAUSSIAN}
            We follow \cite{HIM_Cohen} and \cite{BB_HIM_SYM}
            in demonstrating that (interaction-dressed) Gaussian functions solve
            the corresponding Gross-Pitaevskii equations.
            We exclude the possibility of demixing (symmetry preserving or symmetry broken) 
            in the mean-field solution for the ground state of the trapped mixture.
            To recall, the many-body solution does not exhibit demixing, 
            see the densities (\ref{ONE_B_DENSITIES}).
            While we do not provide a mathematically-rigorous proof
            that other than Gaussians-shaped (symmetry preserving or symmetry broken) solutions
            may occur as the ground-state solution
            of the two-coupled Gross-Pitaevskii equations of the harmonic-interaction model for trapped mixtures,       
            this is not necessary for the proof of Bose-Einstein condensation of the many-body solution.
            Furthermore,            
            the fact that 
            the energy per particle and reduced density matrices per particle
            of the exact many-body ground state and 
            of the Gaussian-shaped
            mean-field solution coincide in the infinite-particle limit
            constitutes in our opinion a rather solid support
            that the Gaussian-shaped orbitals (\ref{MIX_GP_OR}) are indeed 
            the Gross-Pitaevskii ground state. 

\bibitem{Variance} S. Klaiman and O. E. Alon,
                          {\rm Variance as a sensitive probe of correlations}, 
                          Phys. Rev. A {\bf 91}, 063613 (2015).

\bibitem{WAVE_OVLP} S. Klaiman and L. S. Cederbaum,
                                {\rm Overlap of exact and Gross-Pitaevskii wave functions in Bose-Einstein condensates of dilute gases},
                                Phys. Rev. A {\bf 94}, 063648 (2016).

\end{thebibliography}
\end{document}